\pdfoutput=1
\documentclass[11pt, a4paper, onecolumn, copyright, google]{google}

\uselogo{true}

\usepackage{gensymb}
\usepackage{natbib}
\usepackage{xltabular}

\begin{document}

% ======================================================================
%  FRONT MATTER
% ======================================================================
\title{Probabilistic seasonal streamflow forecasting across California's snow-fed basins through agentic discovery}

\author[1]{Ignacio~Lopez-Gomez}
\author[1,2]{Michael P.~Brenner}
\author[1,3]{Tapio~Schneider}

\affil[1]{Google Research, Mountain View, CA, USA}
\affil[2]{Harvard University, Cambridge, MA, USA}
\affil[3]{California Institute of Technology, Pasadena, CA, USA}

\correspondingauthor={ilopezgp@google.com}

\date{\today}

\begin{abstract}
Accurate seasonal runoff forecasts are critical for managing California's reservoirs and its water supply for agriculture and residents. Winter snow accumulation provides a strong source of predictability of snowmelt-based runoff in the spring and summer months, but warming temperatures and progressive hydroclimatic changes are altering its timing and volume. These changes undermine the stationarity assumptions of operational forecasting models and degrade their skill. Here we show that a collaborative agentic framework, combining artificial intelligence systems powered by large language models and tree search with expert human guidance, can discover physics-informed probabilistic forecasting models that improve upon operational systems and capture observed trends in streamflow timing. Guided by human feedback, the agentic system identifies predictor datasets, formulates physically informed features such as snowpack thermal ripeness, and searches over the space of architectures to construct Flowracle, an ensemble forecasting system composed of specialized quantile regression models. Evaluated against operational Bulletin~120 forecasts across 23 snow-fed California watersheds over the 2021--2025 test period, Flowracle reduces statewide quantile errors in April--July runoff forecasts by 36\% in February, 18\% in March, and 27\% in April. The system also shows improved calibration, expands coverage to year-round monthly forecasts across 13 quantiles in all basins, and reduces spring operational volume forecast bias. In addition, the model accurately reproduces observed changes in streamflow timing, demonstrating that expert-guided agentic code search can discover physically grounded, accurate, and climate-resilient models for water availability prediction.
\end{abstract}

\maketitle

% ======================================================================
%  1. INTRODUCTION
% ======================================================================
\section{Introduction}
\label{sec:intro}

Seasonal water supply forecasts guide reservoir operations, agricultural allocations, and flood risk mitigation across the western United States~\citep{Pagano2002, Wood2006}. In California, approximately 60\% of consumptive water originates in the Sierra Nevada~\citep{Rhoades_2018}, where the seasonal snowpack acts as a frozen reservoir whose spring and summer release dictates the timing and volume of usable supply. Accurate predictions of this streamflow, quantified operationally as Full Natural Flow (FNF), are therefore essential for the state's \$60-billion agricultural economy and millions of residents who depend on managed water systems.

For more than half a century, operational seasonal water supply predictions in California have been guided by the Department of Water Resources (DWR) Bulletin~120 (B120) program~\citep{DWR2026}. Bulletin~120 relies primarily on basin-specific multiple linear regression equations calibrated on historical unimpaired streamflow, snow course surveys, and precipitation indices, assuming a peak snowpack date of April 1~\citep{Rizzardo2022}. Although recent efforts have modernized B120 by incorporating airborne snow observations and numerical weather forecasts~\citep{DWR2026_improvements}, its underlying assumptions and statistical relationships, based on historical analogs, are breaking down in a changing climate~\citep{CaliforniaStateAuditor2023, Harrison2016, Milly2008}. Anthropogenic warming shifts precipitation from snow to rain, accelerates the onset of spring melt, and increases the frequency of rain-on-snow events that amplify runoff~\citep{Musselman2018, Siirila-Woodburn2021, Stewart_2005}. B120 forecasts also evaluate each catchment independently, failing to exploit regional spatial coherence and moisture transport across California's mountain watersheds.

Machine learning methods have shown notable skill in hydrologic modeling~\citep{Shen2018, Kratzert2019, Feng2020, Nearing2021}. However, developing operational machine learning systems for seasonal streamflow is constrained by severe data sparsity and non-stationarity. In this regime, expressive architectures may overfit to observational noise or fail to capture extreme drought or flood conditions, the situations when credible forecasts are most consequential~\citep{Dettinger_2011, Swain_2015}. Developing models that generalize reliably requires careful data curation, physics-guided feature engineering, and comprehensive validation, processes that historically demand months of manual experimentation by specialized researchers.

Recent developments in agentic artificial intelligence (AI) systems can reduce this bottleneck through acceleration of the scientific discovery process~\citep{ERA2025, novikov2025, Gottweis2026, lueckmann2026}. Agents powered by large language models (LLMs) can navigate data repositories, synthesize insights from the academic literature, write scientific code, and evaluate hypotheses through computational experimentation. When paired with search algorithms such as Monte Carlo Tree Search (MCTS), agentic systems can probe large combinatorial design spaces of features, architectures, and loss formulations, discovering solutions that exceed human-engineered baselines~\citep{Silver2016, ERA2025}.

Here we present Flowracle, a seasonal streamflow forecasting system for California's snow-fed basins developed through a collaborative agentic framework that outperforms the operational B120 forecasts. 
The framework couples an interactive AI research assistant (Google Antigravity), which conducts multi-modal data discovery, literature synthesis, and produces high-level hypotheses and model implementations, with the Empirical Research Assistant (ERA)~\citep{ERA2025}, which carries out systematic MCTS-driven code mutation and optimization to reduce forecasting errors of the underlying model. The resulting probabilistic forecast system is an ensemble of specialized gradient-boosted decision trees (GBDT) incorporating physics-informed features and seasonally evolving ensemble weighting. Across all 23 Bulletin 120 watersheds over the held-out 2021--2025 test period, Flowracle reduces probabilistic forecast errors, improves tail calibration, and eliminates spring over-prediction bias. It also captures observed elevation-dependent shifts in snowmelt timing, demonstrating its robustness to climate change patterns and rootedness in physical principles.

% ======================================================================
%  2. RESULTS
% ======================================================================
\section{Results}
\label{sec:results}

\subsection{Collaborative agentic development of a streamflow forecasting model}
\label{sec:results_agentic}

\begin{figure}[tbh!]
\centering
\includegraphics[width=0.9\textwidth]{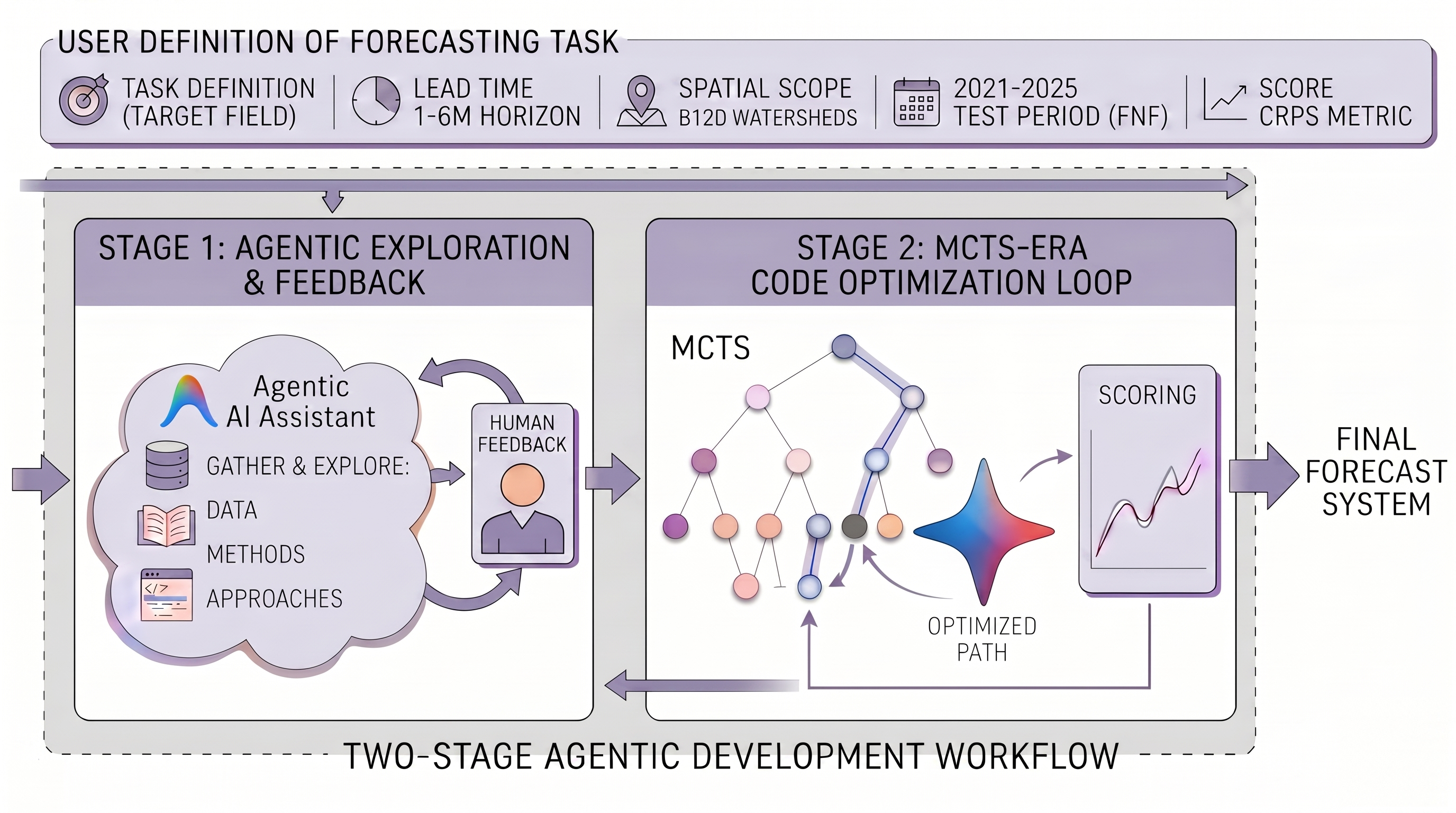}
\caption{Schematic of the two-stage agentic workflow. Stage~1: An interactive AI assistant performs multi-source data discovery, literature synthesis, and baseline model construction under human guidance. Stage~2: The Empirical Research Assistant (ERA) executes MCTS-driven code mutation and optimization over an executable program space. The human researcher iterates between both stages until the final forecasting system achieves the desired level of skill.}\label{fig:workflow}
\end{figure}

Flowracle was developed through a two-stage agentic workflow designed to accelerate scientific discovery (Figure~\ref{fig:workflow}). In Stage~1, Google Antigravity is used to explore candidate predictor datasets, review existing methodologies from the literature, and formulate initial training pipelines. The main human inputs to this stage are high-level task specifications, physical constraints (such as the imposition of non-negative flow), data specifications including operational requirements, and the definition of the evaluation metric to be optimized against, directing the discovery process (Supplementary Table~\ref{tab:prompts}). In Stage~2, preliminary model implementations are submitted to ERA to execute a Monte Carlo Tree Search~\citep{MCTS} over the code space, iteratively mutating feature definitions, model architecture components, regularization schedules, and post-processing steps. Several cycles of the two-stage process may be performed, using the outcomes of each cycle as additional feedback. Human inputs steer the workflow in response to research summaries, modeling suggestions, evaluations, or other agent outputs, functioning much like the oversight a senior researcher provides to a junior colleague (Supplementary Table~\ref{tab:prompts}).

Flowracle frames streamflow prediction as a pooled quantile regression problem across watersheds, predicting monthly Full Natural Flow distributions (13 quantiles) at 1- to 6-month lead times on the first day of each issuance month (Methods). Forecasts use operationally available observations from the preceding six months, combining in situ snow telemetry and streamflow data with gridded meteorological, snowpack, and soil-moisture reanalyses, as well as tropical teleconnection indices. From these inputs, the agentic search constructed physically motivated predictors across regional latitudinal bands and individual catchments, including a proxy of thermal energy available for snowmelt, a soil-moisture priming factor, and the time since peak snowpack conditions. Monotonic constraints are imposed to ensure that higher initial snowpack, soil moisture, precipitation, and streamflow all increase predicted runoff.

For the model architecture, the search converged on an ensemble of three GBDT quantile regressors~\citep{Chen2016} with distinct tree structure, predictor subsets, and target definition: a shallow \textit{Anchor} model ($\mathcal{M}_{\text{anchor}}$) capturing local climatology and bulk moisture accumulation, a deeper \textit{Core} model ($\mathcal{M}_{\text{core}}$) that incorporates nonlinear hydrothermal interaction features, and an \textit{Anomaly} model ($\mathcal{M}_{\text{anom}}$) focused on spatially transferable climatological deviations. Their predictions are blended using monthly varying weights optimized during training. See Methods for the full model description.

The agentic system also introduced non-obvious model components, such as a downward correction applied to the logarithmic quantile predictions before converting back to physical streamflow. The sub-models are trained in log-space to span several orders of magnitude across dry and wet regimes, but finite-sample estimation errors are magnified asymmetrically upon exponentiation, resulting in a positive bias across most quantiles. Subtracting an offset proportional to the cross-validated residual variance from the logarithmic predictions largely eliminated this volume inflation (Methods and Supplementary Section \ref{sec:si_variance_deduction}).

\subsection{Probabilistic streamflow prediction across watersheds}
\label{sec:results_elevation}

\begin{figure}[tbh!]
\centering
\includegraphics[width=\textwidth]{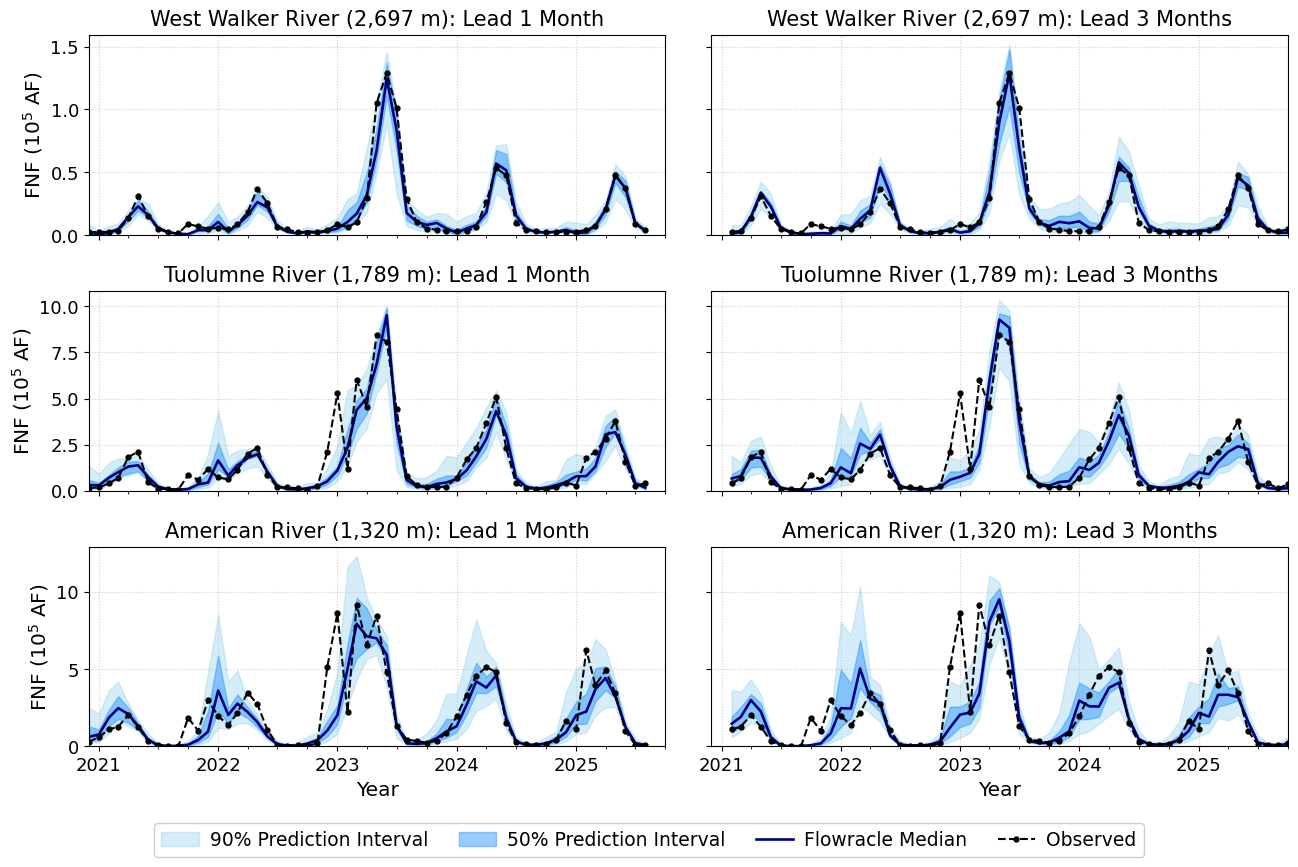}
\caption{Probabilistic forecast time series across a Sierra Nevada elevation transect (2021--2025) for 1- and 3-month lead times. Panels display high-elevation (West Walker River), mid-elevation (Tuolumne River), and low-elevation (American River) catchments. Shaded envelopes denote 90\% and 50\% prediction intervals, with solid and dashed lines indicating median Flowracle forecasts and observations.}\label{fig:runoff_timeseries}
\end{figure}

Flowracle generates year-round monthly forecasts across 13 quantiles for all 23 basins, significantly extending Bulletin 120, which only predicts cumulative April--July volumes at one to three quantile levels. Model selection and cross-validation use data from water years 1981--2020; the final forecasting system is trained on water years 1981--2018, validated on 2019--2020, and evaluated out of sample over 2021--2025 (Methods). To verify whether Flowracle captures hydrologic dynamics across topographical regimes, Figure~\ref{fig:runoff_timeseries} evaluates probabilistic monthly forecast distributions across a Sierra Nevada elevation transect over the 2021--2025 test period: a high-elevation headwater (West Walker River, mean elevation 2,697~m), a mid-elevation snowpack catchment (Tuolumne River, 1,789~m), and a low-elevation rain-snow transition catchment (American River, 1,320~m). Forecast spread expands consistently with lead time and from high to low elevations, consistent with growing meteorological uncertainty and the contrasting predictability of snowmelt and rainfall. Across all three elevation zones, observed monthly flows fall reliably within the 90\% prediction intervals of the forecast, with occasional upper-tail exceedances linked to heavy early-season precipitation below the snow line.

The evaluation period encompasses two contrasting hydrologic extremes. The 2021 drought saw less aggregate annual runoff across the 23 basins than any year in the training record. Conversely, the historic 2023 water year accumulated the largest Sierra Nevada April snowpack in over seven decades~\citep{Marshall_2024}. As shown, the forecasts track both drought recession and atmospheric river runoff surges during these periods, and demonstrate a robust response to extreme inputs. The model also distinguishes elevation-dependent runoff timing, predicting earlier discharge onsets and peaks in watersheds at low elevations. Forecasts for other lead times further demonstrate consistent temporal calibration across horizons (Supplementary Figure~\ref{fig:si_fan_charts}).

\subsection{Operational benchmark against California Bulletin 120}
\label{sec:results_b120}

\begin{figure}[tbh!]
\centering
\includegraphics[width=\textwidth]{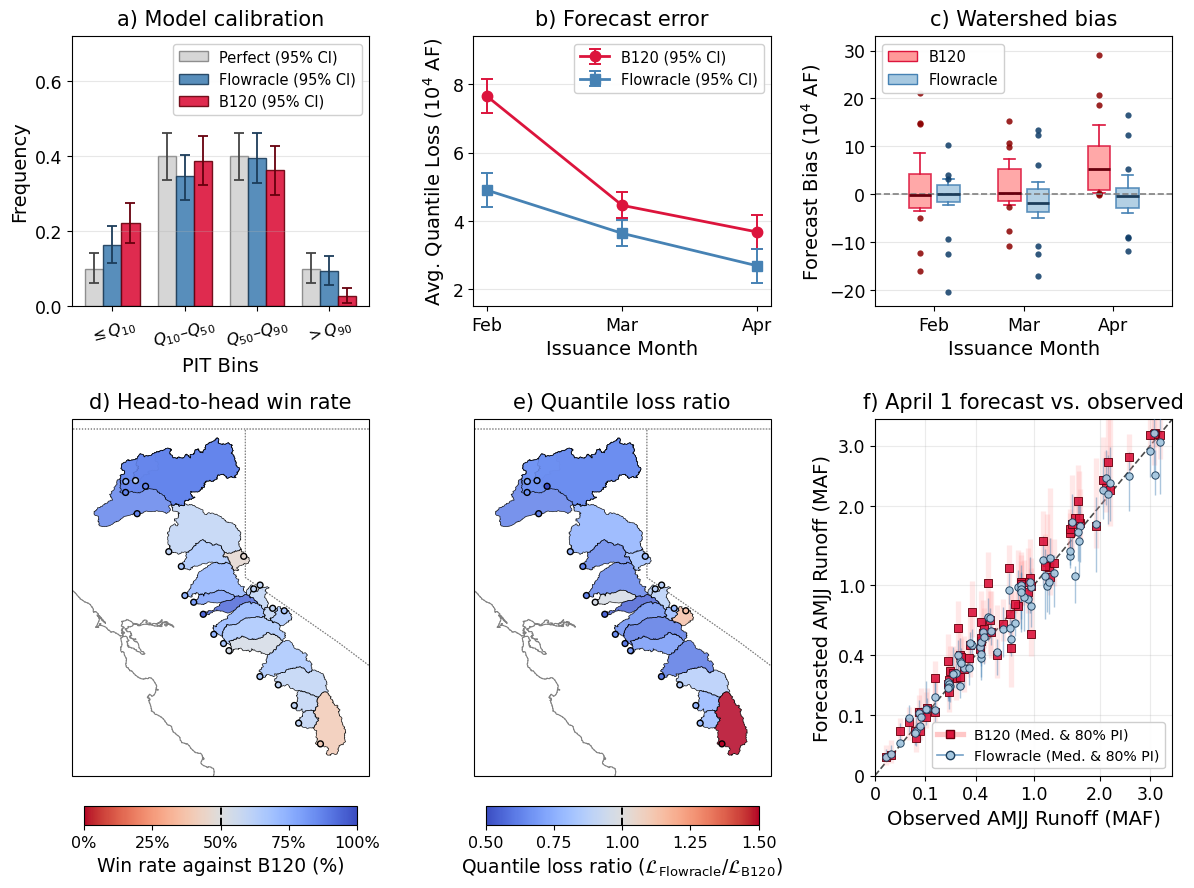}
\caption{Forecast skill of Flowracle and California B120 for February 1, March 1, and April 1 issuances (2021--2025). \textbf{(a)} Probability Integral Transform histogram discretized at 10th, 50th, and 90th percentiles ($N = 225$ paired evaluations). 95\% confidence intervals come from bootstrap resampling for forecasts and binomial consistency envelopes for the reference. \textbf{(b)} Mean quantile loss by issuance month; error bars denote difference-adjusted 95\% paired confidence intervals across watersheds (Supplementary Section~\ref{sec:si_metrics_b120}). \textbf{(c)} Basin distribution of forecast bias. Boxes indicate median and interquartile range, whiskers represent the 10th and 90th percentiles, and points denote outliers. \textbf{(d)} Spatial distribution of win rate. \textbf{(e)} Spatial map of mean quantile loss ratio; values below 1.0 indicate lower loss for Flowracle. \textbf{(f)} April 1 forecasts versus observed April--July runoff (Million Acre-Feet, MAF; $N = 75$); markers denote median forecasts and vertical bars 80\% prediction intervals (PI; $Q_{10}$--$Q_{90}$). B120 publishes extreme deciles for 15 of the 23 watersheds and a median forecast only for the remaining 8. Panels (a) and (f) are restricted to the former 15 watersheds, and in panels (b), (d), and (e) the 8 median-only watersheds are scored for both systems on the median alone (Supplementary Section~\ref{sec:si_metrics_b120}). }\label{fig:quantile_exceedance}
\end{figure}

To assess operational skill, we benchmark Flowracle cumulative April--July (AMJJ) runoff volume forecasts issued in February, March, and April against California Bulletin~120 across 23 snow-fed watersheds (Figure~\ref{fig:quantile_exceedance}).

Probabilistic calibration is evaluated in Figure~\ref{fig:quantile_exceedance}a using a Probability Integral Transform (PIT) histogram across the 15 basins for which B120 publishes extreme decile and median forecasts. B120 exhibits statistically significant tail miscalibration. Observed runoff fell below B120's lower-decile ($Q_{10}$) forecasts more than twice
as often as expected ($22.2\% \pm 5.3\%$, compared to $10\%$). B120's upper-decile ($Q_{90}$) forecasts were also too high, with only $2.7\% \pm 2.0\%$ of observations exceeding them. Flowracle better aligns with the expected frequencies across all probability bins ($16.4\%$ in $\le Q_{10}$, and $9.3\%$ in $> Q_{90}$), resulting in more reliable forecast distributions across hydroclimatic extremes.

Flowracle achieves lower average quantile errors across lead times, as shown in Figure~\ref{fig:quantile_exceedance}b. February model forecasts attain a 36\% error reduction compared to B120, yielding more accurate predictions in 83 of 115 basin-years, or 72.2\%. On March and April issuances, the model achieves 18\% and 27\% error reductions, respectively. Furthermore, the model's February forecast error ($4.9 \times 10^4$~acre-feet (AF)) is comparable to Bulletin~120's March 1 error ($4.5 \times 10^4$~AF), effectively providing March-grade operational accuracy a full month earlier in the snow accumulation season.

Performance across basins is shown in Figures~\ref{fig:quantile_exceedance}d,e. The model achieves win rates (i.e., forecasts with lower quantile error) above 50\% in 21 of the 23 watersheds (monthly spatial maps provided in Supplementary Figure~\ref{fig:si_spatial_win_rates}). Across all forecasts over 2021--2025, the model wins 67.0\% of all head-to-head comparisons. On average, Flowracle achieves lower probabilistic error across most watersheds (Figure~\ref{fig:quantile_exceedance}e), with the largest error reductions concentrated in the southern Cascades and the northern-to-central Sierra Nevada, which feed some of California's primary storage reservoirs (Shasta, Oroville, Folsom, and New Melones) and inter-basin transfer infrastructure (including the California Aqueduct, Delta-Mendota Canal, and Hetch Hetchy system). Skill gains diminish toward the arid southern Sierra, with the model underperforming Bulletin~120 in the Kern River Basin at the southern terminus (win rate $40\%$). This is likely due to the sparse coverage of the southern Sierra Nevada by SNOTEL stations, as well as the use of Airborne Snow Observatory flight data~\citep{Hedrick_2018} by Bulletin 120 during the evaluation period~\citep{DWR2026_improvements}. Flowracle does not use flight data because they have only become available over most basins after the cutoff date of our training period.

Figure~\ref{fig:quantile_exceedance}c examines the basin distribution of forecast bias. For forecasts issued in spring, Bulletin~120 drifts toward positive bias, reaching a median bias of $+5.2 \times 10^4$~AF on April 1 (mean $+7.1 \times 10^4$~AF), over-predicting runoff over the 5-year period in 22 of the 23 basins with individual biases reaching $+29.0 \times 10^4$~AF. Flowracle maintains a centered distribution across all months (April median $-0.5 \times 10^4$~AF, mean $-0.16 \times 10^4$~AF; interquartile range from $-2.9 \times 10^4$ to $+1.3 \times 10^4$~AF). Evaluating statewide volume bias shows that Bulletin~120 tends to over-forecast aggregate runoff, whereas Flowracle largely eliminates this over-allocation risk (Supplementary Figure~\ref{fig:si_statewide_bias}). April-issued forecasts from both systems and observed April--July runoff are compared in Figure~\ref{fig:quantile_exceedance}f. Results for Bulletin~120 confirm its systematic upward bias, with 17\% of observations falling below its lower-decile ($Q_{10}$) forecast and none exceeding its upper decile ($Q_{90}$). Flowracle issues better-aligned forecasts ($R^2 = 0.97$ and $\text{MAE} = 0.10$~MAF, compared to $0.95$ and $0.13$~MAF for Bulletin~120) with well-calibrated 80\% prediction intervals (85\% empirical coverage, with 8\% of observations below $Q_{10}$ and 7\% above $Q_{90}$).

\subsection{Elevation-dependent warming and hydrograph timing fingerprint}
\label{sec:results_edw}

\begin{figure}[tbh!]
\centering
\includegraphics[width=\textwidth]{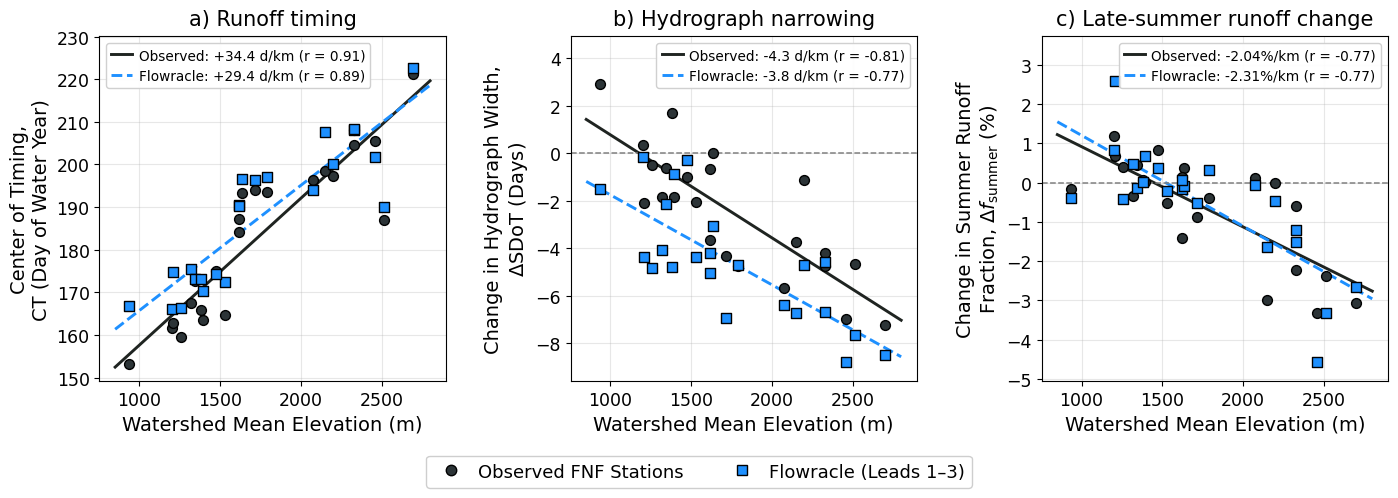}
\caption{Elevation-dependent runoff timing and warming fingerprints across snow-fed California watersheds for observations and forecasts. \textbf{(a)} Runoff timing: annual center of mass of streamflow ($\text{CT}$). \textbf{(b)} Hydrograph narrowing measured in terms of the standard deviation of timing ($\Delta\text{SDoT}$~\citep{Smith_Marshall_2026}; 2021--2025 versus 1981--1990 baseline). \textbf{(c)} Collapse in late-summer runoff: elevation-dependent decline in late-summer (July--September) runoff fraction (2021--2025 versus 1981--1990 baseline). Values are shown for observed gauge records and Flowracle forecasts averaged across 1--3 month lead horizons. In (b) and (c), a stationary baseline is included as a horizontal dashed line for reference.}\label{fig:edw_fingerprint}
\end{figure}

Subject to warming temperatures, snow-fed catchments experience elevation-dependent hydroclimatic shifts that advance runoff timing and compress the seasonal snowmelt window~\citep{Stewart_2005, Smith_Marshall_2026}.
Capturing these shifts, which intensify the operational trade-off between early-season flood releases and summer water availability, is essential for forecasting models to retain skill in a changing climate.
Figure~\ref{fig:edw_fingerprint} examines the ability of Flowracle to capture the timing (Fig.~\ref{fig:edw_fingerprint}a) and observed changes (Fig.~\ref{fig:edw_fingerprint}b, c) in runoff across watersheds as a function of their mean elevation.

Observed unimpaired runoff records reveal an elevation lapse rate of runoff timing of $+34.4$~days/km, shifting the center of timing ($\text{CT}$, the flow-weighted day of the water year; Supplementary Section~\ref{sec:si_metrics_timing}) by up to two months between low- and high-elevation catchments (Figure~\ref{fig:edw_fingerprint}a). Flowracle forecasts track the observed lapse rate across basins ($+29.4~\mathrm{days/km}$). Rising temperatures induce changes in the runoff hydrograph that also vary with elevation (Figure~\ref{fig:edw_fingerprint}b). Comparing the held-out test era (2021--2025) against the historical baseline (1981--1990), observed records display a narrowing lapse rate of $-4.3$~days/km. High-elevation subalpine catchments experience 5 to 7 days of hydrograph compression ($\Delta\text{SDoT}$), whereas lower-elevation basins exhibit little change or slight broadening due to increased winter rainfall as the snow line rises~\citep{Smith_Marshall_2026}. Bulletin~120 predicts bulk seasonal volumes without resolving intra-seasonal timing, implicitly adopting a stationary hydrograph. In contrast, Flowracle predicts a narrowing of $-3.8$~days/km, similar to the observed trend.

This hydrograph narrowing, together with shifts in timing, directly drives a collapse in late-summer water availability (Figure~\ref{fig:edw_fingerprint}c). Observed records show that the fraction of annual runoff occurring in late summer (July--September, $\Delta f_{\text{summer}}$) declined at a rate of $-2.04\%$/km, with high-elevation catchments losing up to 3\% of annual flow volume during that period. Flowracle predicts a decline of $-2.31\%$/km, anticipating late-season water supply deficits across California watersheds compared to its historical training period.

\subsection{Model interpretability and operational feature reliance}
\label{sec:results_shap}

\begin{figure}[tbh!]
\centering
\includegraphics[width=\textwidth]{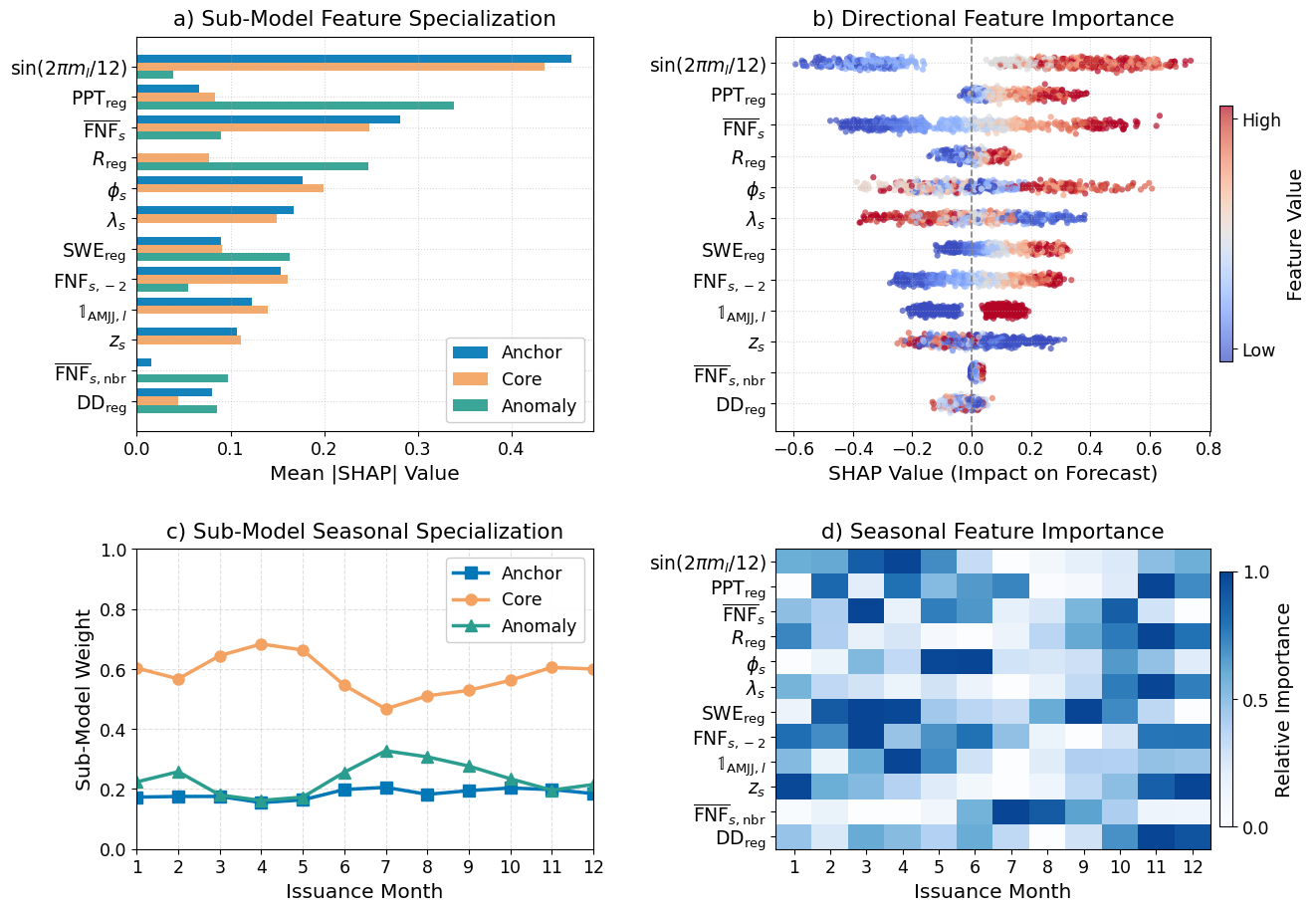}
\caption{Model interpretability and feature attribution diagnostics (2001--2025). \textbf{(a)} Sub-model feature specialization ranked by mean absolute SHAP values for February--April issuances. \textbf{(b)} Directional feature importance from TreeSHAP for February--April issuances. Points denote individual forecasts. \textbf{(c)} Seasonal sub-model blending weights across issuance months. \textbf{(d)} Relative operational feature reliance throughout the year. Feature symbols denote: \textit{calendar harmonics} (target month $m_{l}$ and April--July indicator $\mathbb{1}_{\mathrm{AMJJ}, l}$); \textit{geographic coordinates} (station latitude $\phi_s$, longitude $\lambda_s$, and elevation $z_s$); \textit{antecedent runoff} (local catchment trailing mean $\overline{\mathrm{FNF}}_s$, 2-month lag $\mathrm{FNF}_{s,-2}$, and neighboring basin mean $\overline{\mathrm{FNF}}_{s,\mathrm{nbr}}$); and \textit{hydroclimatic drivers} (regional cumulative precipitation $\mathrm{PPT}_{\mathrm{reg}}$, snow water equivalent $\mathrm{SWE}_{\mathrm{reg}}$, snowpack thermal ripeness $R_{\mathrm{reg}}$, and positive degree-months $\mathrm{DD}_{\mathrm{reg}}$, each aggregated across the four latitudinal bands; Methods, Supplementary Section~\ref{sec:si_metrics_shap}, and Supplementary Table~\ref{tab:features}).}\label{fig:submodel_diagnostics}
\end{figure}

Figure~\ref{fig:submodel_diagnostics} examines the ensemble forecasting mechanism and feature reliance of Flowracle across seasons. The three sub-models exhibit distinct functional specialization, characterized in Figure~\ref{fig:submodel_diagnostics}a using Shapley additive explanations for the most salient features~\citep[TreeSHAP; ][]{lundberg2020}. The directionality of these features is shown in Figure~\ref{fig:submodel_diagnostics}b. Time-of-year indicators and geographic coordinates modulate the response of $\mathcal{M}_{\text{anchor}}$ and $\mathcal{M}_{\text{core}}$, consistent with the strong catchment-specific seasonality of snowmelt-based runoff (Fig.~\ref{fig:submodel_diagnostics}a). In contrast, $\mathcal{M}_{\mathrm{anom}}$ assigns near-zero attribution to $\sin(2\pi m_l / 12)$ because its target already removes each basin's monthly climatological median.

Beyond the seasonal cycle, $\mathcal{M}_{\mathrm{anchor}}$ and $\mathcal{M}_{\mathrm{core}}$ rely heavily on antecedent local streamflow ($\overline{\mathrm{FNF}}_s$ and $\mathrm{FNF}_{s,-2}$) and basin coordinates ($\phi_s$, $\lambda_s$, and $z_s$) to represent catchment scale and baseflow persistence. Higher trailing runoff shifts forecasts upward, whereas more eastern longitudes ($\lambda_s$, rain-shadowed basins) and higher gauge elevations ($z_s$, smaller drainage areas) tend to reduce runoff volume (Figure~\ref{fig:submodel_diagnostics}b). Conversely, $\mathcal{M}_{\mathrm{anom}}$ is governed by dynamic regional processes such as cumulative precipitation ($\mathrm{PPT}_{\mathrm{reg}}$), snowpack thermal ripeness ($R_{\mathrm{reg}}$), snow water equivalent ($\mathrm{SWE}_{\mathrm{reg}}$), neighboring-basin flow ($\overline{\mathrm{FNF}}_{s,\mathrm{nbr}}$), and positive degree-months ($\mathrm{DD}_{\mathrm{reg}}$). While $\mathrm{PPT}_{\mathrm{reg}}$ and $\mathrm{SWE}_{\mathrm{reg}}$ increase predicted runoff monotonically, the relation of some of these features to runoff is complex (e.g., positive degree-months), suggestive of a role as nonlinear modulators (Fig.~\ref{fig:submodel_diagnostics}b).

The adaptive weighting scheme used to produce the ensemble forecasts provides further insights into the seasonal role of each sub-model (Figure~\ref{fig:submodel_diagnostics}c; Eq.~\eqref{eq:shrink}). During the winter and spring accumulation season, the high-capacity $\mathcal{M}_{\mathrm{core}}$ carries the largest weight, as the model must anticipate snowmelt before peak discharge begins. The anomaly sub-model $\mathcal{M}_{\mathrm{anom}}$ gains relevance during the melt and dry season, peaking mid-summer. Finally, the lower-capacity $\mathcal{M}_{\text{anchor}}$ maintains a steady baseline weight throughout the year, anchoring catchment climatology.

Finally, Figure~\ref{fig:submodel_diagnostics}d tracks operational feature reliance across the calendar year. During winter and spring issuances, ensemble predictions are anchored by the seasonal target harmonics ($\sin(2\pi m_l / 12)$ and $\mathbb{1}_{\mathrm{AMJJ}, l}$), regional snowpack ($\mathrm{SWE}_{\mathrm{reg}}$), and local streamflow memory ($\overline{\mathrm{FNF}}_s$ and $\mathrm{FNF}_{s,-2}$). As snowpack recedes during late spring and summer, relative reliance shifts toward watershed latitude ($\phi_s$) and neighboring-basin runoff ($\overline{\mathrm{FNF}}_{s,\mathrm{nbr}}$), tracking the northward and upslope retreat of active melt. In autumn and early winter, before substantial snow accumulation begins, regional precipitation ($\mathrm{PPT}_{\mathrm{reg}}$), thermal predictors ($R_{\mathrm{reg}}$ and $\mathrm{DD}_{\mathrm{reg}}$), and topographic coordinates ($\lambda_s$ and $z_s$) govern early-season forecasts. All predictor features, data sources, and formulations are listed in Supplementary Table~\ref{tab:features}.

% ======================================================================
%  3. DISCUSSION
% ======================================================================
\section{Discussion}
\label{sec:discussion}

Skillful seasonal streamflow forecasting is critical for water supply allocation and flood management in regions where the winter snowpack acts as a reservoir for the dry season, such as the western United States and the Mediterranean~\citep{Wood2006, Dettinger2014, Mankin2015}. Providing reliable predictions of runoff gives reservoir operators essential lead time to adjust flood control space, schedule hydropower generation, and establish carryover conservation pools \citep{Anghileri_2016}. Traditional operational baselines exhibit wide uncertainty at early lead times and biases due to stationary assumptions, limiting the value of currently available forecasts~\citep{Harrison2016, CaliforniaStateAuditor2023}. 

Our results demonstrate that a forecasting system designed using AI agents achieves higher skill than California's operational system across all horizons, reducing average quantile loss across 23 snow-fed catchments by an overall 29\%, and providing February forecasts with operational March-level skill. For water rights allocation and reservoir operations, Flowracle eliminates Bulletin~120's systematic over-prediction on April 1, reducing statewide volume bias from $+1.63$~MAF down to $-0.04$~MAF over the 2021--2025 period. When droughts occur, traditional linear models systematically over-allocate water, creating severe operational deficits when expected snowmelt fails to arrive~\citep{CaliforniaStateAuditor2023}. Reducing the forecast bias directly addresses this operational vulnerability.

Beyond improving forecast accuracy, Flowracle expands operational capability by resolving monthly runoff year-round. Resolving intra-seasonal streamflow allows the system to capture climate change fingerprints that stationary seasonal-volume forecast models cannot represent.
Across California's snow-fed basins, the model reproduces the observed shortening of the snowmelt season and reduction in late-summer runoff, particularly in high-elevation watersheds that have historically sustained late-summer supply. This physical consistency supports the inductive biases introduced during the agentic search process.

This study shows that agentic AI systems for scientific discovery~\citep{ERA2025, novikov2025} can accelerate model development in hydrology when guided by domain experts. Rather than replacing human insight and direction, the agentic framework establishes an effective division of labor. The human scientist defines the physical and operational constraints, data sourcing and curation protocols, and high-level task formulation. The AI systems then explore the design space: searching data repositories and literature insights, engineering meaningful features, and optimizing the model formulation using efficient search algorithms.

Several directions could further enhance the operational value of Flowracle. First, our monthly integration timestep obscures high-frequency sub-monthly hydrologic events, such as rain-on-snow melt pulses and atmospheric river landfall dynamics. Integrating daily hydrometeorological forcing represents a natural extension. Second, purely data-driven architectures remain vulnerable when faced with climatic regimes that lie outside historical observations, even if this vulnerability is reduced through physics-based inductive biases~\citep{Shen2023, Eyring_2024, Beucler_2024, henn2026}. The rapid iteration capacity of agentic workflows offers an additional practical countermeasure, allowing models to be continuously updated and recalibrated as subsequent extreme water years unfold. Finally, coupling this framework with subseasonal weather predictions could further expand lead-time predictability, particularly during the snow accumulation season.

% ======================================================================
%  4. METHODS
% ======================================================================
\section{Methods}
\label{sec:methods}

\subsection{Forecasting task and verification data}
\label{sec:methods_data}

The forecasting target for Flowracle is monthly Full Natural Flow (FNF), estimated by the California Department of Water Resources (DWR) and accessed via the California Data Exchange Center (CDEC). FNF represents the unimpaired runoff volume that would pass a gauging station in the absence of upstream dams, diversions, and other human interventions. We target 23 basins spanning the southern Cascades and Sierra Nevada mountain ranges, encompassing the complete spatial scope of the operational Bulletin~120 program. The study period spans water years 1981 through 2025. Data from 1981 to 2020 are used for training and validation, while data from water years 2021 to 2025 are held out as an out-of-sample test period. The final forecasting system is trained on 1981--2018 (corresponding to the final cross-validation fold, with 2019--2020 held out for validation).

Flowracle is designed to issue probabilistic monthly FNF forecasts one to six months into the future every first of the month. We define $T \in \{1, \ldots, 12\}$ as the calendar issuance month. The lead months $l \in \{1, \ldots, 6\}$ correspond to target calendar months $\big((T + l - 2) \bmod 12\big) + 1$ (for instance, March through August for a March 1 issuance). Therefore, month $T-1$ is the last full elapsed month before forecast issuance. The model provides annual coverage at monthly granularity while evaluating a wider ensemble of quantiles (13) than the operational baseline, which only covers the median and for some basins the extreme deciles.

\subsection{Multi-source predictor suite}
\label{sec:methods_predictors}

Predictor datasets were identified through the agentic discovery process (Section~\ref{sec:results_agentic}) and filtered to ensure operational availability. The resulting suite spans data from station networks, high-resolution gridded reanalyses, and large-scale atmospheric teleconnections (Supplementary Table~\ref{tab:features}).

In situ ground telemetry provides direct observations of snowpack and streamflow. We use past monthly FNF data from CDEC as a predictor. Because monthly FNF estimates incur publication lags, flows for the immediate prior month ($T-1$) are unavailable on the first day of issuance month $T$ and are strictly omitted to eliminate operational data leakage.
Snowpack data are taken from daily and monthly snow water equivalent (SWE) observations from the SNOTEL network~\citep{snotel} alongside CDEC snow sensors (sensor 82).

Hydroclimatic and land-surface conditions are provided by gridded meteorological, snowpack, and drought reanalyses. Spatially distributed monthly precipitation and mean 2-m surface air temperature fields are obtained from the 4-km PRISM reanalysis~\citep{Daly2008}, and gridded monthly-averaged SWE is drawn from the 4-km University of Arizona (UA SWANN) snow reanalysis~\citep{Broxton2016}. Soil moisture and baseflow efficiency are characterized using monthly 4-km Palmer Drought Severity Index (PDSI) fields from gridMET~\citep{Abatzoglou2013, Palmer1965}. To ensure operational availability, the last day of the month is masked from PRISM and UA SWANN monthly aggregates. For PDSI, the last 5 days of the month are masked.

Large-scale atmospheric circulation and tropical convective forcing are indexed through the Real-time Multivariate Madden--Julian Oscillation (RMM) from the Australian Bureau of Meteorology~\citep{Wheeler2004}. Operational availability is ensured by masking the last two days of every month before computing monthly aggregate predictors. Although additional macro-climatic teleconnections, including El Ni\~no--Southern Oscillation and Pacific Decadal Oscillation indices from the National Oceanic and Atmospheric Administration (NOAA), were evaluated during the agentic discovery phase, cross-validation indicated that they induced empirical overfitting across historical training folds and were therefore excluded from the final model.

\subsection{Collaborative agentic discovery and tree search}
\label{sec:methods_agentic_search}

Flowracle was synthesized through collaboration between human researchers, an interactive research assistant (Google Antigravity), and the Empirical Research Assistant (ERA)~\citep{ERA2025}. In Stage~1 (Fig. \ref{fig:workflow}), Antigravity was first prompted to survey the scientific literature and public repositories for modeling
approaches capable of improving upon California's B120 forecasting system. This search included explicit instructions to analyze top-performing models from the U.S. Bureau of Reclamation's Water Supply Forecast Rodeo~\citep{USBR_Rodeo2024}. A key design pattern
extracted from these leading entries was the use of multi-quantile Gradient Boosted Decision Trees
(GBDT) with a quantile error objective, which resulted in significantly better generalization compared
to neural network architectures. Furthermore, the
agent synthesized insights from the hydrological literature to engineer features such as snowpack
ripeness and melt-onset indicators.

Following the data discovery and research stage, we prompted the agentic assistant to explore and formulate a wide range of architectures for the forecast model, given access to the selected datasets, a human-designed multi-source data loader built to prevent temporal data leakage, a scoring function, and the specific training and validation splits (with the test period strictly withheld). Using these inputs and its synthesized domain knowledge, the agent constructed pipelines encompassing data preprocessing, spatiotemporal alignment, feature engineering, model training, and cross-validation for each modeling strategy pursued. Initial implementations, along with their training protocols and low-level design choices, were functional but not yet optimized.

In Stage~2, initial pipelines were transferred to the Empirical Research Assistant (ERA)~\citep{ERA2025} for autonomous optimization. Operating without human intervention in sandboxed containers, ERA used a 500-node Monte Carlo Tree Search (MCTS) to iteratively mutate, execute, and score candidate codebases against a cross-validation loss (Eq.~\eqref{eq:adj} in Section~\ref{sec:methods_metrics}). Code mutations were guided by execution logs, persistent physical rules, and researcher directives targeting regularization, feature engineering, and ensembling (Supplementary Table~\ref{tab:prompts}).

The resulting architectures were evaluated for physical consistency and lead-horizon stability. For example, graph neural networks were discarded because they overfitted at long lead times despite competitive overall training scores. These insights informed iterative prompt refinements with the interactive assistant before returning final candidates to ERA for optimization.

\subsection{Ensemble model architecture}
\label{sec:methods_architecture}

Flowracle formulates monthly runoff prediction as a pooled multi-basin, multi-horizon quantile regression task. For each forecast defined by the target basin $s$, lead horizon $l$, and calendar issuance month $T$, the model predicts unimpaired runoff quantiles $\hat{y}_{s, l, T}^{(\tau)} = f^{(\tau)}(\mathbf{x}_{s, l, T})$ across 13 quantile levels $\tau \in \{0.01, 0.05, 0.1, 0.2, \ldots, 0.9, 0.95, 0.99\}$. The forecast lead horizon $l$ identifies the target calendar month as $m_l = \big((T + l - 2) \bmod 12\big) + 1$.

A single unified model is trained across all 23 basins and 6 lead horizons, allowing cross-basin hydrologic dynamics to inform local predictions. The input feature vector $\mathbf{x}_{s, l, T} \in \mathbb{R}^{55}$ for a given sample $(s, l, T)$ is partitioned into four functional blocks:
\begin{equation}
    \mathbf{x}_{s, l, T} = \big[\mathbf{f}_{\text{region}}(T), \; \mathbf{e}_{\text{lead}}(T, l), \; \boldsymbol{\phi}_s, \; \mathbf{f}_{\text{local}}(s, T)\big],
    \label{eq:feature_vector}
\end{equation}
where $\mathbf{f}_{\text{region}}(T) \in \mathbb{R}^{29}$ captures regional hydroclimatic conditions across latitudinal bands and macro-climate teleconnections; $\mathbf{e}_{\text{lead}}(T, l)  \in \mathbb{R}^{3}$ encodes forecast horizon and target calendar month predictors; $\boldsymbol{\phi}_s = (\phi_s, \lambda_s, z_s)$ denotes the static geographic coordinates (latitude, longitude, elevation) of basin outlet $s$; and $\mathbf{f}_{\text{local}}(s, T) \in \mathbb{R}^{20}$ encompasses dynamic catchment-scale antecedent flows, local reanalyses, and distance-elevation weighted snow telemetry specific to basin $s$ (detailed in Section~\ref{sec:methods_features}).

The operational architecture comprises three specialized XGBoost quantile regression sub-models \citep{Chen2016}, optimized across distinct structural depths, regularization penalties, and feature subsets (Supplementary Table~\ref{tab:submodels}). Two of these sub-models are trained directly on raw log-transformed targets $y_{s,l,T} = \log(1 + \text{FNF}_{s,l,T}/F_s)$ (with $F_s = 10^4$~AF) and retain static basin outlet coordinates $\boldsymbol{\phi}_s$. The anchor model ($\mathcal{M}_{\text{anchor}}$) employs shallow decision trees to provide a stable baseline. In contrast, the core model ($\mathcal{M}_{\text{core}}$) uses deeper trees and includes the snowpack ripeness and melt-timing predictors to capture nonlinear interactions between snow accumulation, temperature, and antecedent soil moisture.

The third component, the anomaly model ($\mathcal{M}_{\text{anom}}$), uses intermediate-depth decision trees trained on deviations ($y_\text{resid}$) from the historical climatological median,
\begin{equation}
    y_{\text{resid}, s, l, T} = y_{s, l, T} - \bar{y}_{s, m_l}.
    \label{eq:resid}
\end{equation}
Here, $\bar{y}_{s, m_l}$ is the sample median of historical log-FNF for basin $s$ in calendar month $m_l$. Input features for $\mathcal{M}_{\text{anom}}$ are similarly centered by their respective monthly climatological medians to learn anomaly relations. By omitting positional coordinates $\boldsymbol{\phi}_s$, the anomaly model enforces transferability across watersheds, while incorporating large-scale atmospheric teleconnections to capture broad circulation anomalies.

\subsection{Physics-guided feature engineering}
\label{sec:methods_features}

High-dimensional spatial fields and telemetry networks are aggregated into physically guided features at regional and catchment scales before being passed to the model (Eq.~\eqref{eq:feature_vector}; detailed algebraic formulations and scaling parameters are provided in Supplementary Section~\ref{sec:si_features} and Supplementary Table~\ref{tab:features}).

\textbf{Regional hydroclimate and teleconnections ($\mathbf{f}_{\text{region}}(T)$):} Regional features are constructed by dividing the spatial domain into four latitudinal bands spanning the Sierra Nevada and southern Cascades: North (N), Mid-North (MN), Mid-South (MS), and South (S). The regional edges are given in Supplementary Section~\ref{sec:si_regional}. For each band $b$, six regional features are extracted from gridded reanalyses within the 6-month pre-issuance window $t \in \{T-6, \dots, T-1\}$: accumulated precipitation ($\text{PPT}_b$), accumulated positive degree-months tracking thermal energy available for snowmelt ($\text{DD}_b$), band-aggregated snow water equivalent ($\text{SWE}_b$), a snowpack ripeness index tracking thermal forcing relative to remaining snowpack ($R_b$), recent monthly SWE rate of change ($\Delta\text{SWE}_b$), and the elapsed time since peak SWE ($\tau_b$; see Supplementary Section~\ref{sec:si_regional}). In addition, $\mathbf{f}_{\text{region}}(T)$ incorporates the monthly mean amplitude and end-of-month polar harmonics of the RMM index, alongside issuance calendar harmonics.

\textbf{Lead horizon predictors ($\mathbf{e}_{\text{lead}}(T, l)$):} Lead-dependent predictors reference the target calendar month $m_l$ through annual Fourier harmonics and an April--July snowmelt season indicator ($\mathbb{1}_{\mathrm{AMJJ}, l}$; Supplementary Section~\ref{sec:si_lead}).

\textbf{Catchment-scale features ($\mathbf{f}_{\text{local}}(s, T)$):} Dynamic predictors local to target basin $s$ integrate nearest-grid reanalyses, distance-weighted ground telemetry, and antecedent streamflow.

For each basin $s$, local accumulated precipitation ($\text{PPT}_{\text{accum}, s}$) and local snowpack ($\text{SWE}_{\mathrm{ua}, s}$) are extracted from the nearest grid cells, together with the local snow-to-precipitation ratio. In situ SWE observations from SNOTEL and CDEC stations are aggregated using three-dimensional inverse-distance weighting across the top $K=32$ nearest sensors to yield log-transformed monthly SWE ($\text{SWE}_{\mathrm{st}, s}; \mathrm{st} \in \{\mathrm{snotel}, \mathrm{cdec}\}$), monthly snowpack rate of change ($\Delta\text{SWE}_{\mathrm{st}, s}$), trailing-month SNOTEL SWE trend ($\Delta_\text{eom}\text{SWE}_{\mathrm{snotel}, s}$), and end-of-month CDEC pillow level and 5-day rate of change ($\text{SWE}_{\mathrm{eom, cdec}, s}$ and $\Delta_{\mathrm{daily}}\text{SWE}_{\mathrm{cdec}, s}$; Supplementary Section~\ref{sec:si_catchment}).

To capture nonlinear runoff modulation by antecedent soil moisture, a primed runoff efficiency predictor ($\eta_{\text{prime}, s}$) is formulated from antecedent PDSI and 6-month accumulated precipitation. The primed runoff efficiency enters the model directly as a predictor and modulates available snowmelt volume and spring melt-onset timing.
Furthermore, to prevent transient snow accumulation from late-winter atmospheric rivers from inflating seasonal streamflow expectations, a late storm damping factor ($D_{\mathrm{late}, s}$) is computed from trailing SNOTEL SWE and applied during spring snowmelt initiation (April 1 issuance).
Autocorrelations in local streamflow are captured by two-month lagged unimpaired flow $\text{FNF}_{s, -2}$ and the 5-month trailing mean $\overline{\text{FNF}}_s$ across operationally available months $[T-6, T-2]$, complemented by regional neighbor flow information ($\mathrm{FNF}_{s,\mathrm{nbr}, -2}$ and $\overline{\mathrm{FNF}}_{s,\mathrm{nbr}}$) from the $K=8$ nearest catchments. Finally, statewide storm coherence ($C_{\mathrm{storm}}$) and state-to-basin snow ratio ($R_{\mathrm{bulk}, s}$) track the spatial uniformity and relative intensity of recent snowpack accumulation (Supplementary Section~\ref{sec:si_catchment}).

\textbf{Monotonic physical constraints.} As a form of inductive bias, monotonic increasing constraints are enforced on 28 of the 55 input features (flagged individually in Supplementary Table~\ref{tab:features}). These directional constraints ensure that higher precipitation, deeper snowpack, wetter antecedent soils, and increased streamflow memory do not reduce predicted runoff volume.

\subsection{Adaptive seasonal ensemble weighting}
\label{sec:methods_weighting}

Sub-model predictions for each quantile $\tau$ are combined through a weighted sum, with weights determined by the RMSE of out-of-fold predictions from a randomly shuffled 4-fold partition of the training samples (distinct from the temporal validation folds of Section~\ref{sec:methods_cv}). Weights for sub-model $k$ are keyed by issuance month $T$ and defined as
\begin{equation}
    {w}_{T}^{(k)} = \alpha \cdot \tfrac{1}{3} + (1-\alpha) \cdot \frac{\big(\text{RMSE}_{T}^{(k)\,\gamma} + \epsilon\big)^{-1}}{\sum_{j} \big(\text{RMSE}_{T}^{(j)\,\gamma} + \epsilon\big)^{-1}},
    \label{eq:shrink}
\end{equation}
with $\alpha = 0.18$, $\gamma = 4.2$, and $\epsilon = 10^{-9}$. The reference climatological median is added to the output of $\mathcal{M}_{\text{anom}}$ before aggregation. The weighted sum $\hat{q}_\tau = \sum_k w_T^{(k)} \hat{q}_\tau^{(k)}$ yields an estimate of the log-quantile, since the sub-models are trained on log-transformed runoff. Before back-transformation, a variance-proportional deduction is subtracted from the blended log-quantile forecasts (Section~\ref{sec:results_agentic} and Supplementary Section~\ref{sec:si_variance_deduction}),
\begin{equation}
    \hat{q}_{\tau, \text{corr}} = \max\!\left(0, \; \hat{q}_\tau - \beta_{\text{bias}} \cdot \sigma^2_{\text{resid}}\right),
    \label{eq:jensen}
\end{equation}
where $\beta_{\text{bias}} = 0.65$ and $\sigma^2_{\text{resid}}$ is the empirical out-of-fold residual variance of the blended ensemble from a 4-fold partition of the training set ($\sigma_{\text{resid}}^2 = 0.0694$ for the 1981--2018 model). Because subtraction happens inside the exponent, it acts as a multiplicative damping factor in physical space. The corrected quantiles $\hat{q}_{\tau, \text{corr}}$ are sorted and linearly interpolated to produce a 100-member probabilistic ensemble, which is back-transformed to physical runoff volume as $\widehat{\text{FNF}} = F_s \cdot (\exp(\hat{q}_{\tau, \text{corr}}) - 1)$.

\subsection{Training and cross-validation}
\label{sec:methods_cv}

To ensure generalization and prevent overfitting, models were cross-validated across three regime-stratified temporal folds: Fold~1 (2011--2014, pre-drought to severe drought), Fold~2 (2015--2017, severe drought to wet extreme), and Fold~3 (2019--2020, wet-to-dry transition). Training strictly used data preceding each validation period, starting from 1981.

The training protocol incorporated target month weighting, quantile-dependent regularization, and stochastic data augmentation. Training samples targeting the critical snowmelt period ($m_l \in \{4, 5, 6, 7\}$) received an explicit sample weight of $w = 2.0$ ($w = 0.5$ for other months) to prioritize peak volume accuracy. To prevent over-shrinkage of extreme quantiles where fewer observations populate terminal tree leaves, the base L2 regularization penalty $\lambda_0$ of each sub-model (Supplementary Table~\ref{tab:submodels}) was scaled by a relaxation factor of $0.6525$ for the outer quantiles ($\tau \le 0.05$ and $\tau \ge 0.95$). In addition, the training set was augmented three-fold across independent epochs using stochastic perturbations to simulate sensor error and calendar-phase variability. Specifically, the augmentation introduced temporal phase interpolation of antecedent snowpack between months $T-2$ and $T-3$, as well as Gaussian noise on historical and target future streamflow, provisional daily CDEC snow pillows, statewide PDSI, and MJO phase and amplitude. Detailed formulations and noise scales are provided in Supplementary Section~\ref{sec:si_augmentation}.

\subsection{Optimization metric}
\label{sec:methods_metrics}

Candidate models were scored using a composite metric blending flow-weighted continuous ranked probability score (CRPS) and cumulative April--July loss. The flow-weighted score is
\begin{equation}
    \text{CRPS}_{\text{FW}} = \tfrac{1}{2} \, \text{CRPS}_{\mathcal{H}} + \tfrac{1}{2} \, \text{CRPS}_{\mathcal{L}},
\end{equation}
where $\mathcal{H}$ (resp., $\mathcal{L}$) denotes validation issuance dates with mean observed flows above (resp., below) the 67th validation set percentile. The normalized composite score for fold $k$ is
\begin{equation}
    \mathcal{L}_k = \frac{1}{2\overline{|y|}_k} \left( \text{CRPS}_{\text{FW}} + \text{CRPS}_{\text{AMJJ}} \right).
\end{equation}
Here $\overline{|y|}_k$ is the mean observed flow magnitude over fold $k$, and $\text{CRPS}_{\text{AMJJ}}$ scores the cumulative April--July volume, averaged over February, March, and April issuances with weights $0.25$, $0.35$, and $0.40$, respectively. To penalize localized underperformance relative to B120, a relative exceedance penalty is computed for each watershed $s$ and issuance month $m \in \{2, 3, 4\}$,
\begin{equation}
    e_{s,m} = \max\!\left(0, \; \frac{\bar{\rho}_{s,m}^{\text{model}} - \bar{\rho}_{s,m}^{\text{B120}}}{\bar{\rho}_{s,m}^{\text{B120}}}\right),
\end{equation}
where $\bar{\rho}_{s,m}^{\text{model}}$ and $\bar{\rho}_{s,m}^{\text{B120}}$ denote the mean April--July quantile losses of the model and B120 for watershed $s$ and issuance month $m$ (defined in Supplementary Section~\ref{sec:si_metrics_b120}). Within each fold $k$, $e_{s,m}$ is averaged across all watershed--month pairs to yield the fold exceedance $\mathcal{E}_{\text{B120}, k}$. The final objective minimized by MCTS is
\begin{equation}
    \mathcal{L}_{\text{adj}} = \bar{\mathcal{L}} + \lambda_{\text{var}} \, \sigma_{\text{fold}} + \lambda_{\text{B120}} \, \mathcal{E}_{\text{B120}},
    \label{eq:adj}
\end{equation}
where $\bar{\mathcal{L}}$ and $\sigma_{\text{fold}}$ are the mean and standard deviation of $\mathcal{L}_k$ across the three temporal folds, $\mathcal{E}_{\text{B120}} = \sum_{k=1}^3 w_{\text{fold},k} \, \mathcal{E}_{\text{B120}, k}$ is the fold-weighted mean exceedance with weights $(0.2, 0.2, 0.6)$ for Folds~1--3, $\lambda_{\text{var}} = 0.10$, and $\lambda_{\text{B120}} = 0.30$. This objective governs model selection only; training relies on the quantile loss for each independent regressor.

% ======================================================================
%  DATA AVAILABILITY
% ======================================================================
\section*{Data Availability}
The datasets used in this study are publicly available from their respective repositories:
Full Natural Flow records, snow sensor data, and Bulletin~120 forecasts from the California Data Exchange Center (\url{https://cdec.water.ca.gov/});
PRISM gridded temperature and precipitation from the PRISM Climate Group (\url{https://prism.oregonstate.edu/});
gridded snow water equivalent from the University of Arizona SWANN reanalysis (\url{https://climate.arizona.edu/data/UA_SWE/});
SNOTEL telemetry from the Natural Resources Conservation Service (\url{https://www.nrcs.usda.gov/wps/portal/wcc/home/});
gridMET Palmer Drought Severity Index from the Climatology Lab (\url{https://www.climatologylab.org/gridmet.html});
and MJO indices from the Australian Bureau of Meteorology (\url{https://www.bom.gov.au/climate/mjo/}).

\section*{Code Availability}
The Flowracle model implementation, data preprocessing notebooks, and evaluation scripts are publicly available under the Apache-2.0 license at \url{https://github.com/google-research/google-research/tree/master/flowracle}.

% ======================================================================
%  ACKNOWLEDGMENTS
% ======================================================================
\section*{Acknowledgments}
We thank Dmitrii Kochkov and Christopher Van Arsdale for their feedback on the manuscript. We gratefully acknowledge the California Data Exchange Center for making Full Natural Flow historical estimates, station data, and Bulletin~120 forecasts publicly available, alongside the University of Arizona for access to gridded snow water equivalent analysis. We also thank the Northwest Alliance for Computational Science and Engineering at Oregon State University for access to the gridded PRISM analysis, the Climatology Lab at UC Merced for maintaining the gridMET data, and the NOAA Physical Sciences Laboratory and Australian Bureau of Meteorology for maintaining an operational repository of macro-climate indices. Finally, we thank the Natural Resources Conservation Service for operating the SNOTEL station network.

% ======================================================================
%  AUTHOR CONTRIBUTIONS
% ======================================================================
\section*{Author Contributions}
I.L.-G., M.P.B., and T.S. conceived the study and designed the research. I.L.-G. developed the forecasting system, conducted the agentic experiments, and performed the data analysis. I.L.-G., M.P.B., and T.S. analyzed the results and wrote the manuscript.

% ======================================================================
%  COMPETING INTERESTS
% ======================================================================
\section*{Competing Interests}
The authors are employees of Alphabet Inc.\ (Google) and own Alphabet stock as part of their standard compensation. The authors declare no other competing interests.

% ======================================================================
%  BIBLIOGRAPHY
% ======================================================================
\bibliography{references}

@article{Pagano2002,
  author  = {Pagano, T. C. and Hartmann, H. C. and Sorooshian, S.},
  title   = {Factors affecting seasonal forecast use in {Arizona} water management: a case study of the 1997--98 {El Ni\~{n}o}},
  journal = {Climate Research},
  year    = {2002},
  volume  = {21},
  pages   = {259--269},
  doi     = {10.3354/cr021259}
}

@article{Rhoades_2018,
author = {Rhoades, Alan M. and Jones, Andrew D. and Ullrich, Paul A.},
title = {The Changing Character of the {California Sierra Nevada} as a Natural Reservoir},
journal = {Geophysical Research Letters},
volume = {45},
number = {23},
pages = {13,008-13,019},
doi = {10.1029/2018GL080308},
year = {2018}
}

@article {Stewart_2005,
      author = "Iris T. Stewart and Daniel R. Cayan and Michael D. Dettinger",
      title = "Changes toward Earlier Streamflow Timing across Western {North} {America}",
      journal = "Journal of Climate",
      year = "2005",
      publisher = "American Meteorological Society",
      address = "Boston MA, USA",
      volume = "18",
      number = "8",
      doi = "10.1175/JCLI3321.1",
      pages=      "1136-1155"
}

@article{Hedrick_2018,
author = {Hedrick, Andrew R. and Marks, Danny and Havens, Scott and Robertson, Mark and Johnson, Micah and Sandusky, Micah and Marshall, Hans-Peter and Kormos, Patrick R. and Bormann, Kat J. and Painter, Thomas H.},
title = {Direct Insertion of {NASA} {Airborne Snow Observatory}-Derived Snow Depth Time Series Into the {iSnobal} Energy Balance Snow Model},
journal = {Water Resources Research},
volume = {54},
number = {10},
pages = {8045-8063},
doi = {10.1029/2018WR023190},
year = {2018}
}

@article{Shen2018,
  author  = {Shen, Chaopeng},
  title   = {A transdisciplinary review of deep learning research and its relevance for water resources scientists},
  journal = {Water Resources Research},
  year    = {2018},
  volume  = {54},
  number  = {11},
  pages   = {8558--8593},
  doi     = {10.1029/2018WR022643}
}

@article{Feng2020,
author = {Feng, Dapeng and Fang, Kuai and Shen, Chaopeng},
title = {Enhancing Streamflow Forecast and Extracting Insights Using Long-Short Term Memory Networks With Data Integration at Continental Scales},
journal = {Water Resources Research},
volume = {56},
number = {9},
pages = {e2019WR026793},
doi = {10.1029/2019WR026793},
year = {2020}
}

@article{Shen2023,
  author  = {Shen, Chaopeng and Appling, Alison P. and Gentine, Pierre and Bandai, Toshiyuki and Gupta, Hoshin and Tartakovsky, Alexandre and Baity-Jesi, Marco and Fenicia, Fabrizio and Kifer, Daniel and Li, Li and Liu, Xiaofeng and Ren, Wei and Zheng, Yi and Harman, Ciaran J. and Clark, Martyn and Farthing, Matthew and Feng, Dapeng and Kumar, Praveen and Aboelyazeed, Doaa and Rahmani, Farshid and Song, Yalan and Beck, Hylke E. and Bindas, Tadd and Dwivedi, Dipankar and Fang, Kuai and H{\"o}ge, Marvin and Rackauckas, Christopher and Mohanty, Binayak and Roy, Tirthankar and Xu, Chonggang and Lawson, Kathryn},
  title   = {Differentiable modelling to unify machine learning and physical models for geosciences},
  journal = {Nature Reviews Earth \& Environment},
  year    = {2023},
  volume  = {4},
  number  = {8},
  pages   = {552--567},
  doi     = {10.1038/s43017-023-00450-9}
}

@article{Wood2006,
  author  = {Wood, Andrew W. and Lettenmaier, Dennis P.},
  title   = {A Test Bed for New Seasonal Hydrologic Forecasting Approaches in the Western {United States}},
  journal = {Bulletin of the American Meteorological Society},
  year    = {2006},
  volume  = {87},
  number  = {12},
  pages   = {1699--1712},
  doi     = {10.1175/BAMS-87-12-1699}
}

@article{Gottweis2026,
   author={Gottweis, Juraj and Weng, Wei-Hung and Daryin, Alexander and Tu, Tao and Sirkovic, Petar and Myaskovsky, Artiom and Glowaty, Grzegorz and Weissenberger, Felix and Orlandi, Alessio and Popovici, Dan and others},
  title     = {Accelerating scientific discovery with {Co-Scientist}},
  journal   = {Nature},
  volume    = {655},
  pages     = {487--496},
  year      = {2026},
  doi       = {10.1038/s41586-026-10644-y}
}

@article{Mankin2015,
  author  = {Mankin, Justin S. and Viviroli, Daniel and Singh, Deepti and Hoekstra, Arjen Y. and Diffenbaugh, Noah S.},
  title   = {The potential for snow to supply human water demand in the present and future},
  journal = {Environmental Research Letters},
  year    = {2015},
  volume  = {10},
  number  = {11},
  pages   = {114016},
  doi     = {10.1088/1748-9326/10/11/114016}
}

@article{Dettinger2014,
  author  = {Dettinger, Michael D. and Cayan, Daniel R.},
  title   = {Drought and the {California Delta}---A Matter of Extremes},
  journal = {San Francisco Estuary and Watershed Science},
  year    = {2014},
  volume  = {12},
  number  = {2},
  doi     = {10.15447/sfews.2014v12iss2art4}
}

@misc{DWR2026,
  author       = {{California Department of Water Resources}},
  title        = {{Bulletin 120}: Water Supply Forecast},
  year         = {2026},
  note = {\url{https://cdec.water.ca.gov/snow/bulletin120/}}
}

@misc{DWR2026_improvements,
  author       = {{California Department of Water Resources}},
  title        = {{DWR} Continues to Improve Forecasting as Spring Heats up in {California}},
  year         = {2026},
  note = {\url{https://water.ca.gov/News/Blog/2026/Mar-2026/DWR-Continues-to-Improve-Forecasting-as-Spring-Heats-up-in-California}}
}

@article{Musselman2018,
  author  = {Musselman, Keith N. and Lehner, Flavio and Ikeda, Kyoko and Clark, Martyn P. and Prein, Andreas F. and Liu, Changhai and Barlage, Mike and Rasmussen, Roy},
  title   = {Projected increases and shifts in rain-on-snow flood risk over western {North America}},
  journal = {Nature Climate Change},
  year    = {2018},
  volume  = {8},
  pages   = {808--812},
  doi     = {10.1038/s41558-018-0236-4}
}

@article{Nearing2021,
  author  = {Nearing, Grey S. and Kratzert, Frederik and Sampson, Alden Keefe and Pelissier, Craig S. and Klotz, Daniel and Frame, Jonathan M. and Prieto, Cristina and Gupta, Hoshin V.},
  title   = {What Role Does Hydrological Science Play in the Age of Machine Learning?},
  journal = {Water Resources Research},
  year    = {2021},
  volume  = {57},
  pages   = {e2020WR028091},
  doi     = {10.1029/2020WR028091}
}

@article{Kratzert2019,
  author  = {Kratzert, Frederik and Klotz, Daniel and Shalev, Guy and Klambauer, G\"{u}nter and Hochreiter, Sepp and Nearing, Grey S.},
  title   = {Towards learning universal, regional, and local hydrological behaviors via machine learning applied to large-sample datasets},
  journal = {Hydrology and Earth System Sciences},
  year    = {2019},
  volume  = {23},
  number  = {12},
  pages   = {5089--5110},
  doi     = {10.5194/hess-23-5089-2019}
}

@inproceedings{Chen2016,
  author    = {Chen, Tianqi and Guestrin, Carlos},
  title     = {{XGBoost}: A Scalable Tree Boosting System},
  booktitle = {Proceedings of the 22nd {ACM SIGKDD} International Conference on Knowledge Discovery and Data Mining},
  year      = {2016},
  pages     = {785--794},
  doi       = {10.1145/2939672.2939785}
}

@article{Broxton2016,
  author  = {Broxton, Patrick D. and Zeng, Xubin and Dawson, Nicholas},
  title   = {Why Do Global Reanalyses and Land Data Assimilation Products Underestimate Snow Water Equivalent?},
  journal = {Journal of Hydrometeorology},
  year    = {2016},
  volume  = {17},
  number  = {11},
  pages   = {2743--2761},
  doi     = {10.1175/JHM-D-16-0056.1}
}

@article{Daly2008,
  author  = {Daly, Christopher and Halbleib, Michael and Smith, Joseph I. and Gibson, Wayne P. and Doggett, Matthew K. and Taylor, George H. and Curtis, Jan and Pasteris, Phillip P.},
  title   = {Physiographically sensitive mapping of climatological temperature and precipitation across the conterminous {United States}},
  journal = {International Journal of Climatology},
  year    = {2008},
  volume  = {28},
  number  = {15},
  pages   = {2031--2064},
  doi     = {10.1002/joc.1688}
}

@article{Wheeler2004,
  author  = {Wheeler, Matthew C. and Hendon, Harry H.},
  title   = {An All-Season Real-Time Multivariate {MJO} Index: Development of an Index for Monitoring and Prediction},
  journal = {Monthly Weather Review},
  year    = {2004},
  volume  = {132},
  number  = {8},
  pages   = {1917--1932},
  doi     = {10.1175/1520-0493(2004)132<1917:AARMMI>2.0.CO;2}
}

@article{ERA2025,
      title={An {AI} system to help scientists write expert-level empirical software}, 
      author={Eser Ayg{\"u}n and Anastasiya Belyaeva and Gheorghe Comanici and Marc Coram and Hao Cui and Jake Garrison and Renee Johnston and Anton Kast and Cory Y. McLean and Peter Norgaard and Zahra Shamsi and David Smalling and James Thompson and Subhashini Venugopalan and Brian P. Williams and Chujun He and Sarah Martinson and Martyna Plomecka and Lai Wei and Yuchen Zhou and Qian-Ze Zhu and Matthew Abraham and Erica Brand and Anna Bulanova and Jeffrey A. Cardille and Chris Co and Scott Ellsworth and Grace Joseph and Malcolm Kane and Ryan Krueger and Johan Kartiwa and Dan Liebling and Jan-Matthis Lueckmann and Paul Raccuglia and Xuefei Julie Wang and Katherine Chou and James Manyika and Yossi Matias and John C. Platt and Lizzie Dorfman and Shibl Mourad and Michael P. Brenner},
      year={2026},
      journal = {Nature},
      volume  = {654},
      pages   = {909--916},
      doi     = {10.1038/s41586-026-10658-6}
}

@article{Milly2008,
  title={Stationarity is dead: Whither water management?},
  author={Milly, Paul C. D. and Betancourt, Julio and Falkenmark, Malin and Hirsch, Robert M and Kundzewicz, Zbigniew W and Lettenmaier, Dennis P and Stouffer, Ronald J},
  journal={Science},
  volume={319},
  number={5863},
  pages={573--574},
  year={2008},
  doi={10.1126/science.1151915},
  publisher={American Association for the Advancement of Science}
}

@article{Siirila-Woodburn2021,
  title={A low-to-no snow future and its impacts on water resources in the western {United States}},
  author={Siirila-Woodburn, Erica R and Rhoades, Alan M and Hatchett, Benjamin J and Huning, Laurie S and Szinai, Julia and Tague, Christina and Nico, Peter S and Feldman, Daniel R and Jones, Andrew D and Collins, William D and Kaatz, Laurna},
  journal={Nature Reviews Earth \& Environment},
  volume={2},
  number={11},
  pages={800--819},
  year={2021},
  doi={10.1038/s43017-021-00219-y},
  publisher={Nature Publishing Group UK, London}
}

@article{Beucler_2024,
author = {Tom Beucler  and Pierre Gentine  and Janni Yuval  and Ankitesh Gupta  and Liran Peng  and Jerry Lin  and Sungduk Yu  and Stephan Rasp  and Fiaz Ahmed  and Paul A. O'Gorman  and J. David Neelin  and Nicholas J. Lutsko  and Michael Pritchard },
title = {Climate-invariant machine learning},
journal = {Science Advances},
volume = {10},
number = {6},
pages = {eadj7250},
year = {2024},
doi = {10.1126/sciadv.adj7250}
}

@article{Anghileri_2016,
author = {Anghileri, D. and Voisin, N. and Castelletti, A. and Pianosi, F. and Nijssen, B. and Lettenmaier, D. P.},
title = {Value of long-term streamflow forecasts to reservoir operations for water supply in snow-dominated river catchments},
journal = {Water Resources Research},
volume = {52},
number = {6},
pages = {4209-4225},
doi = {10.1002/2015WR017864},
year = {2016}
}

@article{Cousineau_2005,
    author =    {Cousineau, Denis },
    journal =   {Tutorials in Quantitative Methods for Psychology},
    publisher = {TQMP},
    title =     {Confidence intervals in within-subject designs: A simpler solution to {Loftus} and {Masson}'s method},
    year =      {2005},
    volume =    {1},
    number =    {1},
    pages =     {42-45},
doi =       {10.20982/tqmp.01.1.p042}
}

@article{Silver2016,
  author  = {Silver, David and Huang, Aja and Maddison, Chris J. and Guez, Arthur and Sifre, Laurent and van den Driessche, George and Schrittwieser, Julian and Antonoglou, Ioannis and Panneershelvam, Veda and Lanctot, Marc and others},
  title   = {Mastering the Game of {Go} with Deep Neural Networks and Tree Search},
  journal = {Nature},
  year    = {2016},
  volume  = {529},
  pages   = {484--489},
  doi     = {10.1038/nature16961}
}

@misc{USBR_Rodeo2024,
  author       = {{U.S. Bureau of Reclamation}},
  title        = {Water Supply Forecast Rodeo},
  year         = {2024},
  howpublished = {DrivenData Competition, \url{https://www.drivendata.org/competitions/group/reclamation-water-supply-forecast/}}
}

@article{Harrison2016,
  author  = {Harrison, Brent and Bales, Roger},
  title   = {Skill Assessment of Water Supply Forecasts for Western {Sierra Nevada} Watersheds},
  journal = {Journal of Hydrologic Engineering},
  year    = {2016},
  volume  = {21},
  number  = {4},
  pages   = {04016002},
  doi     = {10.1061/(ASCE)HE.1943-5584.0001327}
}

@misc{Rizzardo2022,
  author       = {Rizzardo, David},
  title        = {{S2S} Precipitation Forecasts and Snowmelt Runoff Forecasting},
  year         = {2022},
  month        = {May},
  howpublished = {Western States Water Council {S2S} Workshop, San Diego, CA},
  note         = {\url{https://westernstateswater.org/wp-content/uploads/2022/03/DRizzardo-WaterSupplyForecasting-S2S-workshop-May2022-San-Diego-final.pdf}}
}

@techreport{CaliforniaStateAuditor2023,
  author      = {{California State Auditor}},
  title       = {{Department of Water Resources}: Its Forecasts Do Not Adequately Account for Climate Change and Its Reasons for Some Reservoir Releases Are Unclear},
  institution = {California State Auditor},
  year        = {2023},
  month       = {May},
  type        = {Audit Report},
  number      = {2022-106},
  note        = {\url{https://information.auditor.ca.gov/pdfs/reports/2022-106.pdf}}
}

@ARTICLE{MCTS,
  author={Browne, Cameron B. and Powley, Edward and Whitehouse, Daniel and Lucas, Simon M. and Cowling, Peter I. and Rohlfshagen, Philipp and Tavener, Stephen and Perez, Diego and Samothrakis, Spyridon and Colton, Simon},
  journal={IEEE Transactions on Computational Intelligence and AI in Games}, 
  title={A Survey of {Monte Carlo} Tree Search Methods}, 
  year={2012},
  volume={4},
  number={1},
  pages={1-43},
doi={10.1109/TCIAIG.2012.2186810}}

@article{snotel,
author = {Serreze, Mark C. and Clark, Martyn P. and Armstrong, Richard L. and McGinnis, David A. and Pulwarty, Roger S.},
title = {Characteristics of the western {United States} snowpack from snowpack telemetry ({SNOTEL}) data},
journal = {Water Resources Research},
volume = {35},
number = {7},
pages = {2145-2160},
doi = {10.1029/1999WR900090},
year = {1999}
}

@article{Swain2018,
  author    = {Swain, Daniel L. and Langenbrunner, Baird and Neelin, J. David and Hall, Alex},
  title     = {Increasing precipitation volatility in twenty-first-century {California}},
  journal   = {Nature Climate Change},
  volume    = {8},
  pages     = {427--433},
  year      = {2018},
  doi       = {10.1038/s41558-018-0140-y}
}

@article{Eyring_2024,
 title={Pushing the frontiers in climate modelling and analysis with machine learning},
 volume={14},
 DOI={10.1038/s41558-024-02095-y},
 number={9}, journal={Nature Climate Change},
 author={Eyring, Veronika and Collins, William D. and Gentine, Pierre and Barnes, Elizabeth A. and Barreiro, Marcelo and Beucler, Tom and Bocquet, Marc and Bretherton, Christopher S. and Christensen, Hannah M. and Dagon, Katherine and Gagne, David John and Hall, David and Hammerling, Dorit and Hoyer, Stephan and Iglesias-Suarez, Fernando and Lopez-Gomez, Ignacio and McGraw, Marie C. and Meehl, Gerald A. and Molina, Maria J. and Monteleoni, Claire and Mueller, Juliane and Pritchard, Michael S. and Rolnick, David and Runge, Jakob and Stier, Philip and Watt-Meyer, Oliver and Weigel, Katja and Yu, Rose and Zanna, Laure},
 year={2024},
 pages={916--928} }

@article{Marshall_2024,
author = {Adrienne M. Marshall  and John T. Abatzoglou  and Stefan Rahimi  and Dennis P. Lettenmaier  and Alex Hall },
title = {California's 2023 snow deluge: Contextualizing an extreme snow year against future climate change},
journal = {Proceedings of the National Academy of Sciences},
volume = {121},
number = {20},
pages = {e2320600121},
year = {2024},
doi = {10.1073/pnas.2320600121}
}

@misc{novikov2025,
      title={{AlphaEvolve}: A coding agent for scientific and algorithmic discovery}, 
      author={Alexander Novikov and Ng{\^a}n V{\~u} and Marvin Eisenberger and Emilien Dupont and Po-Sen Huang and Adam Zsolt Wagner and Sergey Shirobokov and Borislav Kozlovskii and Francisco J. R. Ruiz and Abbas Mehrabian and M. Pawan Kumar and Abigail See and Swarat Chaudhuri and George Holland and Alex Davies and Sebastian Nowozin and Pushmeet Kohli and Matej Balog},
      year={2025},
      eprint={2506.13131},
      archivePrefix={arXiv},
      primaryClass={cs.AI},
      note    = {\url{https://arxiv.org/abs/2506.13131}}
}

@misc{lueckmann2026,
      title={Discovering Mechanistic Models of Neural Activity: System Identification in an in Silico Zebrafish}, 
      author={Jan-Matthis Lueckmann and Viren Jain and Micha{\l} Januszewski},
      year={2026},
      eprint={2602.04492},
      archivePrefix={arXiv},
      primaryClass={q-bio.NC},
      note    = {\url{https://arxiv.org/abs/2602.04492}}
}

@misc{henn2026,
      title={{AIMIP Phase 1}: systematic evaluations of {AI} weather and climate models}, 
      author={Brian Henn and Christopher S. Bretherton and Nikolay Koldunov and Christian Lessig and Maria J. Molina and Troy Arcomano and Oliver Watt-Meyer and Guillaume Couairon and Renu Singh and Robert Brunstein and Yana Hasson and Antonia Jost and Noah Brenowitz and Peter Manshausen and Nathaniel Cresswell-Clay and Dale Durran and Kyle Joseph Chen Hall and Janni Yuval and Dmitrii Kochkov and Stephan Hoyer and Ignacio Lopez-Gomez},
      year={2026},
      eprint={2605.06944},
      archivePrefix={arXiv},
      primaryClass={physics.ao-ph},
      note    = {\url{https://arxiv.org/abs/2605.06944}}
}

@article{Smith_Marshall_2026,
  author  = {Smith, Steven M. and Marshall, Adrienne M.},
  title   = {Narrowing streamflow distribution can alter water allocation timing and quantity},
  journal = {Nature Water},
  year    = {2026},
  volume = {4},
  pages = {742--751},
  doi     = {10.1038/s44221-026-00639-4}
}

@article{Swain_2015,
author = {Swain, Daniel L.},
title = {A tale of two {California} droughts: Lessons amidst record warmth and dryness in a region of complex physical and human geography},
journal = {Geophysical Research Letters},
volume = {42},
number = {22},
pages = {9999-10,003},
doi = {10.1002/2015GL066628},
year = {2015}
}

@Article{Dettinger_2011,
AUTHOR = {Dettinger, Michael D. and Ralph, Fred Martin and Das, Tapash and Neiman, Paul J. and Cayan, Daniel R.},
TITLE = {Atmospheric Rivers, Floods and the Water Resources of {California}},
JOURNAL = {Water},
VOLUME = {3},
YEAR = {2011},
NUMBER = {2},
PAGES = {445--478},
DOI = {10.3390/w3020445}
}

@article{lundberg2020,
  title={From local explanations to global understanding with explainable {AI} for trees},
  author={Lundberg, Scott M. and Erion, Gabriel and Chen, Hugh and DeGrave, Alex and Prutkin, Jordan M. and Nair, Bala and Katz, Ronit and Himmelfarb, Jonathan and Bansal, Nisha and Lee, Su-In},
  journal={Nature Machine Intelligence},
  volume={2},
  number={1},
  pages={56--67},
  year={2020},
  doi={10.1038/s42256-019-0138-9}
}

@article{Abatzoglou2013,
  author  = {Abatzoglou, John T.},
title = {Development of gridded surface meteorological data for ecological applications and modelling},
journal = {International Journal of Climatology},
volume = {33},
number = {1},
pages = {121-131},
doi = {10.1002/joc.3413},
year = {2013}
}

@techreport{Palmer1965,
  author      = {Palmer, Wayne C.},
  title       = {Meteorological Drought},
  institution = {U.S. Department of Commerce, Weather Bureau},
  number      = {45},
  year        = {1965},
  type = {Research Paper}
}

% ======================================================================
%  SUPPLEMENTARY INFORMATION
% ======================================================================
\clearpage
\appendix
\section*{Supplementary Information}
\renewcommand{\thefigure}{S\arabic{figure}}
\renewcommand{\theHfigure}{S\arabic{figure}}
\setcounter{figure}{0}
\renewcommand{\thetable}{S\arabic{table}}
\renewcommand{\theHtable}{S\arabic{table}}
\setcounter{table}{0}
\renewcommand{\theequation}{S\arabic{equation}}
\renewcommand{\theHequation}{S\arabic{equation}}
\setcounter{equation}{0}
\setcounter{section}{0}
\renewcommand{\thesection}{\arabic{section}}
\renewcommand{\theHsection}{S\arabic{section}}
\titleformat{\section}
{\large\bfseries\headingfont}
{Supplementary Section \thesection:}
{0.5em}
{#1}
[]

\section{Detailed model description}
\label{sec:si_features}

This section details the formulation of the predictors, as well as data augmentation and post-processing techniques used by Flowracle. Supplementary Table~\ref{tab:features} summarizes all features.

\subsection{Regional hydroclimate and teleconnection predictors \texorpdfstring{($\mathbf{f}_{\text{region}}(T)$)}{(f\_region(T))}}
\label{sec:si_regional}
Regional features are constructed by dividing the spatial domain into four latitudinal bands: North (N; 39.5--42\,$^\circ$N), Mid-North (MN; 38--39.5\,$^\circ$N), Mid-South (MS; 36.5--38\,$^\circ$N), and South (S; 32.5--36.5\,$^\circ$N). The zonal extent of all bands is $124^\circ\text{W}$ to $115^\circ\text{W}$. For each band $b$, six regional features are constructed from the 6-month pre-issuance window $t \in \{T-6, \dots, T-1\}$, with individual aggregations specified below. Teleconnection and calendar features are also defined here. Throughout, $\bar{X}_{b,t}$ denotes spatial averaging over band $b$ and indexing at pre-issuance month $t$:
\begin{itemize}
    \item Cumulative precipitation ($\text{PPT}_b$): PRISM precipitation is accumulated and log-transformed,
    \begin{equation}
        \text{PPT}_b = \log\!\left(1 + \sum_{t=T-6}^{T-1} \bar{P}_{b,t} / P_s\right),
        \label{eq:ppt}
    \end{equation}
    where $\bar{P}_{b,t}$ is band-averaged monthly precipitation and $P_s = 60$~mm.

    \item Cumulative positive degree-months ($\text{DD}_b$): Thermal energy available for snowmelt is estimated as
    \begin{equation}
        \text{DD}_b = \sum_{t=T-6}^{T-1} \max(0, \bar{T}_{b,t}) / T_s,
        \label{eq:dd}
    \end{equation}
    where $\bar{T}_{b,t}$ is the band-averaged PRISM 2-m mean temperature (in $^\circ$C), and $T_s=30^\circ$C.

    \item Snow water equivalent ($\text{SWE}_b$): UA~SWANN SWE from month $T-1$ is band-aggregated and log-transformed,
    \begin{equation}
        \text{SWE}_b = \log\!\left(1 + \overline{\text{SWE}}_{b,T-1} / S_s\right),
        \label{eq:swe}
    \end{equation}
    with snowpack scale $S_s = 450$~mm.

    \item Snowpack thermal ripeness index ($R_b$): A dimensionless index tracking cumulative thermal forcing relative to remaining snowpack,
    \begin{equation}
        R_b = \min\!\left(\text{DD}_b / (\overline{\text{SWE}}_{b,T-1} / S_R + 1), \; 10\right),
        \label{eq:ripeness}
    \end{equation}
    with scale parameter $S_R = 100$~mm.

    \item SWE rate of change ($\Delta\text{SWE}_b$): The regional SWE change between months $T-1$ and $T-2$,
    \begin{equation}
        \Delta\text{SWE}_b = (\overline{\text{SWE}}_{b,T-1} - \overline{\text{SWE}}_{b, T-2}) / S_{V},
        \label{eq:swe_vel}
    \end{equation}
    with scale $S_V = 50$~mm.

    \item Time since peak SWE ($\tau_b$): The elapsed time (in months) from the peak band-averaged SWE within the 6-month pre-issuance window,
    \begin{equation}
        \tau_b = (T - 1) - \operatorname*{arg\,max}_{t \in \{T-6,\ldots,T-1\}} \overline{\text{SWE}}_{b,t}.
        \label{eq:time_since_peak}
    \end{equation}

    \item Issuance calendar harmonics: Annual harmonics of issuance month $T \in \{1, \dots, 12\}$,
    \begin{equation}
        \mathbf{h}_{\mathrm{iss}}(T) = \left[\sin\!\left(\frac{2\pi T}{12}\right), \; \cos\!\left(\frac{2\pi T}{12}\right)\right].
        \label{eq:iss_harmonics}
    \end{equation}

    \item Madden--Julian Oscillation amplitude and polar harmonics: Tropical convective forcing is indexed through the Real-time Multivariate MJO indices (RMM1 and RMM2) from the Australian Bureau of Meteorology. The feature vector $\mathbf{h}_{\mathrm{mjo}}$ includes the end-of-month polar phase angle $\theta_{\mathrm{mjo}}$ and the monthly mean amplitude $A_{\mathrm{mjo}}$,
    \begin{equation}
        \theta_{\mathrm{mjo}} = \operatorname{arctan2}(\mathrm{RMM2}, \; \mathrm{RMM1}),
        \label{eq:mjo_phase}
    \end{equation}
    \begin{equation}
        A_{\mathrm{mjo}} = \frac{1}{D - 1}\sum_{d=1}^{D - 1} \sqrt{\mathrm{RMM1}_d^2 + \mathrm{RMM2}_d^2},
        \label{eq:mjo_amp}
    \end{equation}
    \begin{equation}
        \mathbf{h}_{\mathrm{mjo}} = \left[A_{\mathrm{mjo}}, \; \sin(\theta_{\mathrm{mjo}}), \; \cos(\theta_{\mathrm{mjo}})\right],
        \label{eq:mjo_harmonics}
    \end{equation}
    where $(\mathrm{RMM1}_d, \mathrm{RMM2}_d)$ denote the daily indices on day $d$ of month $T-1$ (with the final two days masked for operational availability), and $\mathrm{RMM1}$ and $\mathrm{RMM2}$ are their averages over the trailing 7 operationally available days ($d \in \{D-8, \dots, D-2\}$).
\end{itemize}

\subsection{Lead horizon predictors \texorpdfstring{($\mathbf{e}_{\text{lead}}(T, l)$)}{(e\_lead(T, l))}}
\label{sec:si_lead}
For any lead horizon $l \in \{1, \dots, 6\}$, the target calendar month $m_l = ((T + l - 2) \bmod 12) + 1$ provides the temporal reference for lead-dependent predictors:
\begin{itemize}
    \item Annual target-month Fourier harmonics: $\sin(2\pi m_l / 12)$ and $\cos(2\pi m_l / 12)$.
    \item Snowmelt season indicator: $\mathbb{1}_{\mathrm{AMJJ}, l} = \mathbb{1}_{m_l \in \{4, 5, 6, 7\}}$, distinguishing whether the forecast target falls within the April--July period.
\end{itemize}
These features form the lead encoding vector
\begin{equation}
    \mathbf{e}_{\text{lead}}(T, l) = \left[\sin\!\left(\frac{2\pi m_l}{12}\right), \; \cos\!\left(\frac{2\pi m_l}{12}\right), \; \mathbb{1}_{\mathrm{AMJJ}, l}\right].
    \label{eq:lead_encoding}
\end{equation}

\subsection{Catchment-scale predictors \texorpdfstring{($\mathbf{f}_{\text{local}}(s, T)$)}{(f\_local(s, T))}}
\label{sec:si_catchment}
Dynamic predictors local to target basin $s$ integrate nearest-grid reanalyses, distance-weighted in situ measurements, and antecedent streamflow. Local features are indexed by basin $s$ and pre-issuance month $t$, $X_{s,t}$. Many scaling factors are common to regional and local features and defined above.

Reanalysis-derived features are taken from the nearest grid cell. In situ SNOTEL and CDEC snow pillow observations (Section~\ref{sec:methods_predictors}) are aggregated using an inverse-distance weighting across the $K=32$ sensors closest to target basin $s$,
\begin{equation}
    d_{s, k}^2 = (\phi_k - \phi_s)^2 + (\lambda_k - \lambda_s)^2 + \left(\frac{z_k - z_s}{d_e}\right)^2,
    \label{eq:sensor_dist}
\end{equation}
where $(\phi_s, \lambda_s, z_s)$ and $(\phi_k, \lambda_k, z_k)$ denote the coordinates (\degree) and elevation (m) of the basin outlet and reporting sensor $k$, respectively, and $d_e = 720$~m is a scaling factor. Aggregation weights are assigned as $w_{s,k} \propto 1 / (d_{s,k}^2 + 0.1)$. This yields basin-specific monthly time series of SNOTEL SWE ($\text{SWE}_{\mathrm{snotel}, s, t}$), cumulative SNOTEL precipitation since the start of the water year ($\text{PPT}_{\mathrm{snotel}, s, t}$), and CDEC snow pillow SWE ($\text{SWE}_{\mathrm{cdec}, s, t}$). Daily measurements covering the last 5 days of the trailing month $T-1$ from CDEC pillow records ($\text{SWE}^{(d)}_{\mathrm{cdec}, s}$) and SNOTEL ($\text{SWE}^{(d)}_{\mathrm{snotel}, s}$) are also considered, where $d \in \{D-4, \dots, D\}$ indexes the day and $D$ is the last day of the month.

These station aggregations are then used to construct model predictors as detailed below. All catchment-scale predictors are used for all basins, with the exception of predictors based on CDEC snow pillows (i.e., not SNOTEL). These are only used for basins with gauges south of 37\degree N, to remedy the sparse SNOTEL coverage of the southern Sierra Nevada.

The catchment-scale predictors for basin $s$ comprise:
\begin{itemize}
    \item Local cumulative precipitation: Nearest-grid accumulated PRISM precipitation $P_{s,t}$, log-transformed,
    \begin{equation}
        \text{PPT}_{\text{accum}, s} = \log\!\left(1 + \sum_{t=T-6}^{T-1} P_{s,t} / P_s\right).
        \label{eq:local_ppt}
    \end{equation}
    \item Local gridded SWE: Snowpack from the nearest UA~SWANN grid cell at month $T-1$,
    \begin{equation}
        \text{SWE}_{\mathrm{ua}, s} = \log\!\left(1 + \text{SWE}_{\mathrm{ua}, s, T-1} / S_s\right).
        \label{eq:local_ua_swe}
    \end{equation}
    \item Local snow-to-precipitation ratio: Ratio of local snowpack to 6-month accumulated precipitation,
    \begin{equation}
        f_{\mathrm{snow}, s} = \log\!\left(1 + \frac{\text{SWE}_{\mathrm{ua}, s, T-1}}{\sum_{t=T-6}^{T-1} P_{s,t} + 1}\right).
        \label{eq:local_snow_frac}
    \end{equation}

    \item Primed runoff efficiency ($\eta_{\text{prime}, s}$): Activation function tracking how soil saturation affects the conversion from precipitation to runoff,
    \begin{equation}
        \eta_{\text{prime}, s} = \frac{1}{1 + \exp(-0.5 \cdot \text{Priming}_s)},
        \label{eq:primed_efficiency}
    \end{equation}
    where $\text{Priming}_s$ is an intermediate variable combining antecedent PDSI and cumulative precipitation,
    \begin{equation}
        \text{Priming}_s = 0.6 \cdot \text{PDSI}_{T-1} + 0.4 \cdot \left(\frac{1}{P_s}\sum_{t=T-6}^{T-1} P_{s,t} - 5.0\right).
        \label{eq:priming}
    \end{equation}
    Saturated conditions yield runoff efficiency values close to 1, whereas dry soil pushes the efficiency to 0. Besides entering the model as a predictor, $\eta_{\text{prime}, s}$ modulates other predictors as detailed below, acting as a nonlinear activation function.
    \item Station-aggregated SWE: antecedent SNOTEL and CDEC snowpack,
    \begin{equation}
        \text{SWE}_{\mathrm{snotel}, s} = \log\!\left(1 + \text{SWE}_{\mathrm{snotel}, s, T-1} / S_s\right),
        \label{eq:snotel_swe}
    \end{equation}
    \begin{equation}
        \text{SWE}_{\mathrm{cdec}, s} = \log\!\left(1 + \text{SWE}_{\mathrm{cdec}, s, T-1} / S_s\right).
        \label{eq:cdec_swe}
    \end{equation}
    \item End-of-month CDEC pillow SWE: Last available day snowpack level from CDEC pillows,
    \begin{equation}
        \text{SWE}_{\mathrm{eom, cdec}, s} = \log\!\left(1 + \text{SWE}^{(D)}_{\mathrm{cdec}, s} / S_s\right).
        \label{eq:daily_cdec_swe}
    \end{equation}
    \item Station-aggregated SWE monthly rate of change: SNOTEL and CDEC change between months $T-1$ and $T-2$,
    \begin{equation}
        \Delta\text{SWE}_{\mathrm{snotel}, s} = (\text{SWE}_{\mathrm{snotel}, s, T-1} - \text{SWE}_{\mathrm{snotel}, s, T-2}) / S_V,
        \label{eq:snotel_vel}
    \end{equation}
    \begin{equation}
        \Delta\text{SWE}_{\mathrm{cdec}, s} = (\text{SWE}_{\mathrm{cdec}, s, T-1} - \text{SWE}_{\mathrm{cdec}, s, T-2}) / S_V.
        \label{eq:cdec_vel}
    \end{equation}

    \item Trailing-month SNOTEL SWE trend: Difference between end-of-month and monthly mean SNOTEL SWE,
    \begin{equation}
        \Delta_\text{eom}\text{SWE}_{\mathrm{snotel}, s} = (\text{SWE}^{(D)}_{\mathrm{snotel}, s} - \text{SWE}_{\mathrm{snotel}, s, T-1}) / S_V.
        \label{eq:eom_diff}
    \end{equation}

    \item End-of-month CDEC snow rate of change: Trailing 5-day snowpack rate of change from daily CDEC pillows,
    \begin{equation}
        \Delta_{\mathrm{daily}}\text{SWE}_{\mathrm{cdec}, s} = (\text{SWE}^{(D)}_{\mathrm{cdec}, s} - \text{SWE}^{(D-4)}_{\mathrm{cdec}, s}) / 5.0.
        \label{eq:daily_cdec_vel}
    \end{equation}

    \item SNOTEL snow-to-precipitation ratio: Ratio of SNOTEL snowpack to cumulative SNOTEL precipitation,
    \begin{equation}
        f_{\mathrm{sn\_snow}, s} = \log\!\left(1 + \frac{\text{SWE}_{\mathrm{snotel}, s, T-1}}{\text{PPT}_{\mathrm{snotel}, s, T-1} + 1}\right).
        \label{eq:snotel_snow_frac}
    \end{equation}

    \item Primed available snowmelt volume: A proxy for effective snowpack volume available for runoff after replenishing soil moisture,
    \begin{equation}
        V_{\mathrm{melt}, s} = \log\!\left(1 + \frac{\text{SWE}_{\mathrm{snotel}, s, T-1}}{S_s} \cdot \eta_{\text{prime}, s}\right).
        \label{eq:primed_melt_vol}
    \end{equation}

    \item Primed melt-onset pulse: Proxy for trailing-month melt available for runoff during the melt season,
    \begin{equation}
        M_{\mathrm{onset}, s} = \Delta_\text{eom}\text{SWE}_{\mathrm{snotel}, s} \cdot \mathbb{1}_{\mathrm{melt}, T} \cdot \eta_{\text{prime}, s},
        \label{eq:primed_melt_timing}
    \end{equation}
    where $\mathbb{1}_{\mathrm{melt}, T} = \mathbb{1}_{T \in \{4,\dots,8\}}$ acts as a seasonal melt gate.

    \item Local antecedent streamflow: Log-transformed runoff volume at month $T-2$, as well as average runoff across months $[T-6, T-2]$,
    \begin{equation}
        \text{FNF}_{s, -2} = \log\!\left(1 + \text{FNF}_{s, T-2} / F_s\right),
        \label{eq:fnf_m2}
    \end{equation}
    \begin{equation}
        \overline{\text{FNF}}_s = \log\!\left(1 + \frac{1}{5\,F_s}\sum_{t=T-6}^{T-2} \text{FNF}_{s,t}\right),
        \label{eq:fnf_mem}
    \end{equation}
    with streamflow scale $F_s = 10^4$\,AF (month $T-1$ is excluded; see Section~\ref{sec:methods_data}).
    \item Antecedent regional streamflow: Inverse-distance-weighted antecedent runoff from the $K=8$ nearest neighboring catchments, indexed over the same time periods as the local streamflow,
    \begin{equation}
        \mathrm{FNF}_{s, \mathrm{nbr}, -2} = \log\!\left(1 + \mathrm{FNF}_{s, \mathrm{nbr}, T-2} / F_s\right),
        \label{eq:nbr_fnf_m2}
    \end{equation}
    \begin{equation}
        \overline{\mathrm{FNF}}_{s, \mathrm{nbr}} = \log\!\left(1 + \frac{1}{5\,F_s}\sum_{t=T-6}^{T-2} \mathrm{FNF}_{s, \mathrm{nbr}, t}\right).
        \label{eq:nbr_fnf_avg}
    \end{equation}

    \item Cross-basin storm coherence: Dimensionless indicator capturing the uniformity of recent snow accumulation across basins,
    \begin{equation}
        C_{\mathrm{storm}} = \frac{1}{1 + \sigma(\mathbf{v}_{\mathrm{sn}}) / (|\bar{v}_{\mathrm{sn}}| + 1.0)},
        \label{eq:storm_coherence}
    \end{equation}
    where $\mathbf{v}_{\mathrm{sn}} = [\text{SWE}_{\mathrm{snotel}, s, T-1} - \text{SWE}_{\mathrm{snotel}, s, T-2}]_{s=1}^S$ denotes the vector of monthly SNOTEL SWE changes across basins, $\bar{v}_{\mathrm{sn}}$ is its spatial mean, and $\sigma(\mathbf{v}_{\mathrm{sn}})$ its standard deviation.

    \item State-to-basin snow ratio: Ratio of the statewide mean positive snow accumulation to the local station snowpack,
    \begin{equation}
        R_{\mathrm{bulk}, s} = \frac{\frac{1}{S}\sum_{s'=1}^S \max(0, \; \text{SWE}_{\mathrm{snotel}, s', T-1} - \text{SWE}_{\mathrm{snotel}, s', T-2})}{\text{SWE}_{\mathrm{snotel}, s, T-1} + 1.0}.
        \label{eq:storm_bulk_ratio}
    \end{equation}
\end{itemize}

\subsection{Late storm damping for April issuances \texorpdfstring{($D_{\mathrm{late}, s}$)}{(D\_late,s)}}
\label{sec:si_surge}
Flowracle introduces a damping term for April 1 issuances based on the fraction of accumulated snow that precipitates late in the season. The term aims to capture differences in expected future runoff from snowfall during the cold early winter season and the warmer months. It enters the model as a modulator of SWE predictors rather than a direct predictor.

The late storm damping factor $D_{\mathrm{late}, s}$ is evaluated from SNOTEL station SWE. First, the ratio of recent March snow accumulation relative to the trailing winter snowpack is computed as
\begin{equation}
    \text{SurgeRatio}_s = \operatorname{clip}\!\left(\frac{\max(0, \; \text{SWE}_{\mathrm{snotel}, s, T-1} - \text{SWE}_{\mathrm{snotel}, s, T-2})}{\overline{\text{SWE}}_{s,3\text{mo}} + S_{\mathrm{base}}}, \; 0, \; 3\right),
    \label{eq:surge_ratio}
\end{equation}
where $\overline{\text{SWE}}_{s,3\text{mo}} = \frac{1}{3}\sum_{k=1}^3 \text{SWE}_{\mathrm{snotel}, s, T-k}$ is the trailing 3-month mean SNOTEL SWE, and $S_{\mathrm{base}} = 25$~mm. The damping factor is formulated as
\begin{equation}
    D_{\mathrm{late}, s} = \frac{1}{1 + \beta_{\mathrm{surge}} \cdot \text{SurgeRatio}_s \cdot \mathbb{1}_{T = \text{April}}},
    \label{eq:surge_damping}
\end{equation}
with $\beta_{\mathrm{surge}} = 0.75$. The indicator function $\mathbb{1}_{T = \text{April}}$ strictly restricts damping to April 1 issuances. This damping term selectively attenuates snowpack volume and rate of change predictors:
\begin{align}
    \text{SWE}_{\mathrm{snotel}, s, T-1}^* &= \text{SWE}_{\mathrm{snotel}, s, T-1} \cdot D_{\mathrm{late}, s}, \label{eq:damped_sn} \\
    \text{SWE}_{\mathrm{cdec}, s, T-1}^* &= \text{SWE}_{\mathrm{cdec}, s, T-1} \cdot D_{\mathrm{late}, s}, \label{eq:damped_ca} \\
    \text{SWE}_{\mathrm{ua}, s, T-1}^* &= \text{SWE}_{\mathrm{ua}, s, T-1} \cdot D_{\mathrm{late}, s}, \label{eq:damped_ua} \\
    \Delta_{\mathrm{daily}}\text{SWE}_{\mathrm{cdec}, s}^* &= \Delta_{\mathrm{daily}}\text{SWE}_{\mathrm{cdec}, s} \cdot D_{\mathrm{late}, s}. \label{eq:damped_daily_vel}
\end{align}
These damped quantities replace their raw counterparts in the station SNOTEL SWE (Eq.~\eqref{eq:snotel_swe}), CDEC SWE (Eq.~\eqref{eq:cdec_swe}), end-of-month CDEC SWE rate of change (Eq.~\eqref{eq:daily_cdec_vel}), local gridded SWE (Eq.~\eqref{eq:local_ua_swe}), local snow fraction (Eq.~\eqref{eq:local_snow_frac}), SNOTEL snow fraction (Eq.~\eqref{eq:snotel_snow_frac}), and primed available melt volume (Eq.~\eqref{eq:primed_melt_vol}) predictors.

\subsection{Data augmentation during training}
\label{sec:si_augmentation}
To prevent overfitting to observational noise and sensitivity to calendar month boundaries, training batches are augmented across three epochs using stochastic jitter before model fitting. During each training epoch, the following random perturbations are applied to the input features and target runoff:
\begin{itemize}
    \item \textit{Temporal phase shift of snowpack changes}: Stochastically interpolating antecedent snowpack between months $T-2$ and $T-3$ with random weight $\alpha \sim \mathcal{U}(0, 0.15)$ acts as temporal phase augmentation,
    \begin{equation}
        \text{SWE}_{T-2} \leftarrow (1 - \alpha) \cdot \text{SWE}_{T-2} + \alpha \cdot \text{SWE}_{T-3}.
        \label{eq:jitter_lag}
    \end{equation}
    Because California's snow accumulation is driven by episodic atmospheric rivers, calendar month boundaries arbitrarily bin individual storms into either month $T-3$ or $T-2$ based on differences of only a few days. Blending the antecedent snowpack prevents models from overfitting when computing snowpack change rates ($\Delta\text{SWE}_b$, Eq.~\eqref{eq:swe_vel}; $\Delta\text{SWE}_{\mathrm{snotel}, s}$, Eq.~\eqref{eq:snotel_vel}; $\Delta\text{SWE}_{\mathrm{cdec}, s}$, Eq.~\eqref{eq:cdec_vel}) and the trailing winter baseline ($\text{SurgeRatio}_s$, Eq.~\eqref{eq:surge_ratio}), while leaving the month $T-1$ state unperturbed to preserve the snowpack available for seasonal runoff.
    \item \textit{Daily snow sensor noise}: To reflect provisional measurement fluctuations in snow pillows, end-of-month daily CDEC records ($\text{SWE}^{(d)}_{\mathrm{cdec}, s}$) are perturbed with multiplicative Gaussian noise,
    \begin{equation}
        \text{SWE}^{(d)}_{\mathrm{cdec}, s} \leftarrow \text{SWE}^{(d)}_{\mathrm{cdec}, s} \cdot (1 + \epsilon_{\mathrm{daily}}), \quad \epsilon_{\mathrm{daily}} \sim \mathcal{N}(0, 0.07^2),
        \label{eq:jitter_cdec}
    \end{equation}
    propagating into end-of-month SWE ($\text{SWE}_{\mathrm{eom, cdec}, s}$, Eq.~\eqref{eq:daily_cdec_swe}) and rate of change ($\Delta_{\mathrm{daily}}\text{SWE}_{\mathrm{cdec}, s}$, Eq.~\eqref{eq:daily_cdec_vel}).
    \item \textit{Soil moisture perturbation}: Additive Gaussian noise is applied to the statewide Palmer Drought Severity Index,
    \begin{equation}
        \text{PDSI}_{T-1} \leftarrow \text{PDSI}_{T-1} + \epsilon_{\mathrm{pdsi}}, \quad \epsilon_{\mathrm{pdsi}} \sim \mathcal{N}(0, 0.08^2),
        \label{eq:jitter_pdsi}
    \end{equation}
    which propagates directly into soil priming ($\text{Priming}_s$, Eq.~\eqref{eq:priming}), primed runoff efficiency ($\eta_{\text{prime}, s}$, Eq.~\eqref{eq:primed_efficiency}), primed melt volume ($V_{\mathrm{melt}, s}$, Eq.~\eqref{eq:primed_melt_vol}), and primed melt pulse ($M_{\mathrm{onset}, s}$, Eq.~\eqref{eq:primed_melt_timing}).
    \item \textit{Antecedent streamflow noise}: Historical runoff ($\text{FNF}$) receives multiplicative Gaussian jitter,
    \begin{equation}
        \text{FNF} \leftarrow \text{FNF} \cdot (1 + \epsilon_{\mathrm{fnf}}), \quad \epsilon_{\mathrm{fnf}} \sim \mathcal{N}(0, 0.015^2),
        \label{eq:jitter_fnf}
    \end{equation}
    affecting local antecedent streamflow ($\text{FNF}_{s, -2}$, Eq.~\eqref{eq:fnf_m2}; $\overline{\text{FNF}}_s$, Eq.~\eqref{eq:fnf_mem}) and neighbor streamflow ($\mathrm{FNF}_{s, \mathrm{nbr}, -2}$, Eq.~\eqref{eq:nbr_fnf_m2}; $\overline{\mathrm{FNF}}_{s, \mathrm{nbr}}$, Eq.~\eqref{eq:nbr_fnf_avg}).
    \item \textit{Madden--Julian Oscillation teleconnection jitter}: Gaussian angular jitter is added to the phase angle ($\theta_{\mathrm{mjo}} \leftarrow \theta_{\mathrm{mjo}} + \epsilon_\theta$, with $\epsilon_\theta \sim \mathcal{N}(0, 0.05^2)$~rad), and multiplicative Gaussian noise is applied to the amplitude ($A_{\mathrm{mjo}} \leftarrow A_{\mathrm{mjo}} \cdot (1 + \epsilon_A)$, with $\epsilon_A \sim \mathcal{N}(0, 0.10^2)$) to prevent overfitting to historical MJO phases (Eq.~\eqref{eq:mjo_harmonics}).
    \item \textit{Target runoff perturbation}: During training, the target unimpaired runoff volume $\text{FNF}_{s,l,T}$ is modulated with multiplicative Gaussian noise ($\text{FNF}_{s,l,T} \leftarrow \text{FNF}_{s,l,T} \cdot (1 + \epsilon_y)$, with $\epsilon_y \sim \mathcal{N}(0, 0.012^2)$).
\end{itemize}

\subsection{Log-space variance correction}
\label{sec:si_variance_deduction}
Although exact population quantiles are equivariant under monotonic transformations, finite-sample estimators carry a variance $\sigma^2_{\text{est}}$. Log-quantile estimates from Flowracle may therefore introduce positive or negative sampling errors. Transformation of these estimates through exponentiation amplifies positive errors more than negative ones, introducing a positive bias in physical space. Minimizing expected quantile loss in physical space (Eq.~\eqref{eq:si_pinball}) therefore requires correcting for this asymmetry.

To this end, Flowracle introduces a correction proportional to a proxy of the estimator variance, $\sigma^2_{\text{resid}} \approx \sigma^2_{\text{est}}$. The proxy is obtained by training the three GBDT sub-models using a squared error loss on a 4-fold split of the augmented training set (Supplementary Section \ref{sec:si_augmentation}). $\sigma^2_{\text{resid}}$ is then computed as the out-of-fold mean squared error of the model predictions, and the scaling parameter $\beta_\mathrm{bias}$ in Eq.~\eqref{eq:jensen} is obtained during the Monte Carlo Tree Search process. Subtracting the correction across all quantiles lowers average quantile loss across $\tau \in \{0.10, 0.50, 0.90\}$ by $2.0\%$ for single-month forecasts and $3.1\%$ for cumulative April--July forecasts, while reducing median volume bias by $63.8\%$.

\clearpage

\section{Verification metrics}
\label{sec:si_metrics}

This section defines the diagnostics used in Sections~\ref{sec:results_b120} and~\ref{sec:results_edw}.

\subsection{Probabilistic verification against Bulletin 120}
\label{sec:si_metrics_b120}

\textit{Evaluation design.} We compare the cumulative April--July (AMJJ) runoff volume from Flowracle to the B120 predictions. Forecasts are verified for the 23 snow-fed watersheds that B120 covers, over the held-out water years 2021--2025, at the three issuance dates $T \in \{$February 1, March 1, April 1$\}$, giving $N = 23 \times 5 = 115$ forecasts per issuance and $345$ in total. For an issuance at month $T$, the AMJJ volume is assembled from Flowracle forecasts for lead horizons $l$ whose target months $m_l$ fall in April--July,
\begin{equation}
  \hat{V}^{(j)}_{s,T} = \sum_{l \,:\, m_l \in \{4,\dots,7\}} \hat{y}^{(j)}_{s,T,l},
  \label{eq:si_amjj_sum}
\end{equation}
so that February, March, and April issuances aggregate lead months $3$--$6$, $2$--$5$, and $1$--$4$, respectively. Summing ordered ensemble members across months $m_l \in \{4,\dots,7\}$ amounts to a comonotonic assumption that reflects the strong month-to-month rank persistence of spring snowmelt in California (mean historical Spearman correlation $\bar{\rho}_S = 0.87$ between adjacent months), driven by the large interannual variability of precipitation in the state~\citep{Dettinger_2011, Swain2018}.

\textit{Quantile loss.} For a predicted quantile $\hat{q}_\tau$ at level $\tau$ and observation $y$, the quantile (pinball) loss is
\begin{equation}
  \rho_\tau(y, \hat{q}_\tau) = \max\!\left\{\tau\,(y - \hat{q}_\tau),\; (\tau - 1)(y - \hat{q}_\tau)\right\}.
  \label{eq:si_pinball}
\end{equation}
B120 is disseminated at quantile levels $\tau \in \{0.1, 0.5, 0.9\}$ for most basins. For these basins, the benchmark score is the average pinball loss over that triplet,
\begin{equation}
  \rho(y) = \tfrac{1}{3}\left[\rho_{0.1}(y, \hat{q}_{0.1}) + \rho_{0.5}(y, \hat{q}_{0.5}) + \rho_{0.9}(y, \hat{q}_{0.9})\right].
  \label{eq:si_loss3q}
\end{equation}
For watersheds where B120 only issues the median, all models are scored on the median alone with Eq.~\eqref{eq:si_pinball}. This convention compares the two systems on the richest forecast product they share at each site. Scoring both systems on the median alone at those 8 sites avoids penalizing B120 for intervals it does not issue, making the reported skill gains conservative.

\textit{Skill summaries.} Writing $\rho^{\text{model}}_{s,T,k}$ and $\rho^{\text{B120}}_{s,T,k}$ for the average quantile losses at watershed $s$, issuance $T$, and water year $k$, the head-to-head win rate is the fraction of individual forecasts on which Flowracle attains a lower error,
\begin{equation}
  W = \frac{1}{|\mathcal{T}|}\sum_{(s,T,k) \in \mathcal{T}} \mathbb{1}\!\left[\rho^{\text{model}}_{s,T,k} < \rho^{\text{B120}}_{s,T,k}\right],
  \label{eq:si_winrate}
\end{equation}
evaluated per issuance month ($|\mathcal{T}| = 115$), per watershed across all horizons ($|\mathcal{T}| = 15$; Figure~\ref{fig:quantile_exceedance}d), and pooled ($|\mathcal{T}| = 345$). The loss ratio mapped in Figure~\ref{fig:quantile_exceedance}e is $\overline{\rho}^{\text{model}}_s / \overline{\rho}^{\text{B120}}_s$, each term averaged over the 15 trials at watershed $s$. Reported error reductions are $1 - \overline{\rho}^{\text{model}} / \overline{\rho}^{\text{B120}}$ with averages taken over all 23 watersheds, and the error bars in Figure~\ref{fig:quantile_exceedance}b represent difference-adjusted 95\% paired confidence intervals across the $N=23$ watersheds~\citep{Cousineau_2005}.

\textit{Probability integral transform.} Calibration is assessed by binning each observation into the four intervals $\{y \le Q_{10}\}$, $\{Q_{10} < y \le Q_{50}\}$, $\{Q_{50} < y \le Q_{90}\}$ and $\{y > Q_{90}\}$, whose probabilities under a perfectly calibrated forecast are $(0.10, 0.40, 0.40, 0.10)$. This diagnostic requires published outer quantiles from both systems and is therefore restricted to the 15 watersheds where these are available from B120, giving $N = 225$ paired evaluations. The consistency envelope for observations represents the $95\%$ binomial interval $[F^{-1}(0.025; N, p), F^{-1}(0.975; N, p)]/N$ for each nominal probability $p$, which for the outer bins spans $6.2\%$ to $14.2\%$ and for the inner bins $33.8\%$ to $46.2\%$. Error bars for forecasts represent $95\%$ confidence intervals over $2{,}000$ bootstrapped samples.

\begin{figure}[tbh!]
\centering
\includegraphics[width=0.78\textwidth]{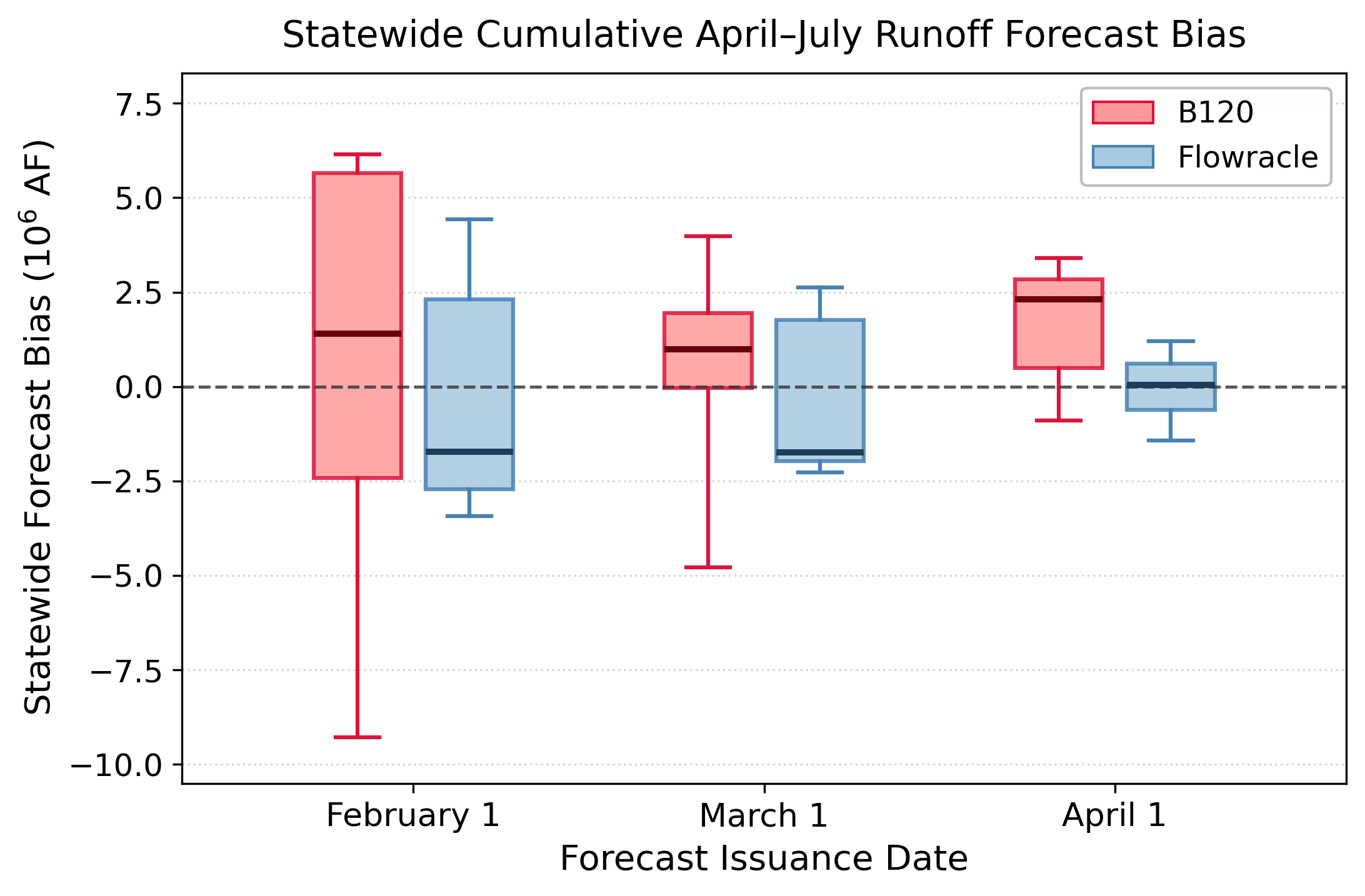}
\caption{Statewide cumulative April--July runoff forecast bias for February, March, and April issuances (2021--2025). Boxplots depict the median, interquartile range, and full range across the five test water years for Bulletin~120 and Flowracle.}\label{fig:si_statewide_bias}
\end{figure}

\textit{Volume bias and scatter plots.} Forecast bias is computed as the median forecast minus the observed volume, $\hat{q}_{0.5} - y$ (Figure~\ref{fig:quantile_exceedance}c), and summed across watersheds to give a statewide total (Supplementary Figure~\ref{fig:si_statewide_bias}). In Figure~\ref{fig:quantile_exceedance}f, the coefficient of determination $R^2 = 1 - \sum (y - \hat{q}_{0.5})^2 / \sum (y - \bar{y})^2$ and the mean absolute error $N^{-1} \sum |\hat{q}_{0.5} - y|$ are computed on median forecasts over the $N = 75$ April 1 evaluations (15 watersheds $\times$ 5 water years) for which B120 publishes all three quantiles. Over the same evaluations, the lower- and upper-tail exceedance rates are the fractions of observations with $y < Q_{10}$ and $y > Q_{90}$, respectively; their complement is the empirical coverage of the nominal 80\% prediction interval.

\subsection{Runoff timing diagnostics}
\label{sec:si_metrics_timing}

Let $q_{s,m}$ denote the monthly unimpaired runoff volume at watershed $s$ in month $m = 1, \dots, 12$ of a water year beginning 1 October, and let $d_m$ be the mid-month day of that water year, e.g., $d_2 = 45$. The center of timing is the flow-weighted mean day~\citep{Stewart_2005},
\begin{equation}
  \mathrm{CT}_s = \frac{\sum_{m} d_m\, q_{s,m}}{\sum_{m} q_{s,m}},
  \label{eq:si_ct}
\end{equation}
and the standard deviation of timing~\citep{Smith_Marshall_2026} is the corresponding flow-weighted dispersion,
\begin{equation}
  \mathrm{SDoT}_s = \left[\frac{\sum_{m} \left(d_m - \mathrm{CT}_s\right)^2 q_{s,m}}{\sum_{m} q_{s,m}}\right]^{1/2},
  \label{eq:si_sdot}
\end{equation}
which measures the length of the runoff season.

The late-summer runoff fraction is the share of the annual volume delivered in July--September,
\begin{equation}
  f_{\text{summer},s} =\frac{\sum_{m = 10}^{12} q_{s,m}}{\sum_{m} q_{s,m}}.
  \label{eq:si_fsummer}
\end{equation}
Each metric is computed per water year and then averaged over the historical baseline decade 1981--1990 and over the held-out period 2021--2025. $\Delta\mathrm{SDoT}$ and $\Delta f_{\text{summer}}$ denote the difference between the two period means. The corresponding modeled quantities are constructed from median forecasts: for each target month of each test water year, the value predicted at lead $l$ is inserted into a 12-month hydrograph. Equations~\eqref{eq:si_ct}--\eqref{eq:si_fsummer} are then applied relative to the observed 1981--1990 baseline, and results are averaged over horizons $l \in \{1, 2, 3\}$. Each diagnostic is regressed on watershed mean elevation, and the slope is reported per kilometer.

\subsection{Feature attribution and interpretability diagnostics}
\label{sec:si_metrics_shap}
\textit{TreeSHAP attributions and regional aggregation.} Feature attributions in Section~\ref{sec:results_shap} (Figure~\ref{fig:submodel_diagnostics}) are evaluated over 2001--2025 across all watersheds and lead horizons using exact TreeSHAP~\citep{lundberg2020} on the median forecast of each sub-model. Predictors excluded from a sub-model's feature subset (Supplementary Table~\ref{tab:submodels}) receive zero attribution. To quantify the combined contribution of each regional feature across the four latitudinal bands, band-level SHAP values are summed per sample by Shapley additivity,
\begin{equation}
  \phi_{n, g_{\mathrm{reg}}}^{(k)} = \sum_{b \in \{\mathrm{N, MN, MS, S}\}} \phi_{n, g_b}^{(k)},
  \label{eq:si_shap_reg}
\end{equation}
consolidating the 24 band-specific inputs into 6 unified regional predictors and yielding 37 distinct physical predictors from the 55 raw features.

\textit{Sub-model specialization and directional importance.} The importance of predictor $j$ in sub-model $k$ is defined as the mean absolute attribution $I_j^{(k)} = \frac{1}{N}\sum_{n=1}^N |\phi_{n,j}^{(k)}|$ over $N = 800$ randomly sampled February--April basin--lead instances. Figure~\ref{fig:submodel_diagnostics}a displays the top 12 predictors ranked by $\max_k I_j^{(k)}$. In Figure~\ref{fig:submodel_diagnostics}b, ensemble attributions for each sample are obtained by blending sub-model SHAP values with the ensemble weights $\bar{w}_{\mathrm{spring}}^{(k)} = \frac{1}{3} \sum_{T=2}^{4} w_T^{(k)}$ (Eq.~\eqref{eq:shrink}),
\begin{equation}
  \phi_{n,j}^{\text{ens}} = \sum_{k} \bar{w}_{\text{spring}}^{(k)}\, \phi_{n,j}^{(k)},
  \label{eq:si_shap_ens}
\end{equation}
with points colored by the 5th--95th percentile-normalized predictor value.

\textit{Seasonal feature reliance.} To resolve seasonal shifts in predictor reliance across the water year (Figure~\ref{fig:submodel_diagnostics}d), mean absolute attributions $I_{j,T}^{(k)}$ are evaluated separately for each issuance month ($N_T = 250$ randomly sampled basin--lead instances per month) and combined using the corresponding monthly ensemble weights, $I_{j,T}^{\text{ens}} = \sum_k {w}_T^{(k)} I_{j,T}^{(k)}$. Each predictor row is min--max normalized across the 12 months to $[0, 1]$.

\clearpage

\begingroup
\small
\setlength{\tabcolsep}{4.5pt}
\renewcommand{\arraystretch}{1.08}
\begin{xltabular}{\textwidth}{|>{\raggedright\arraybackslash}p{3.2cm}|>{\raggedright\arraybackslash}p{2.0cm}|>{\raggedright\arraybackslash}X|}
\caption{Representative human prompts and guidelines provided during the collaborative agentic development workflow.}\label{tab:prompts}\\
\hline
\textbf{Directive Type} & \textbf{Target System} & \textbf{Researcher Prompt or Directive} \\
\hline
\endfirsthead

\multicolumn{3}{l}{\small\textit{Supplementary Table~\thetable, continued from previous page}} \\[4pt]
\hline
\textbf{Directive Type} & \textbf{Target System} & \textbf{Researcher Prompt or Directive} \\
\hline
\endhead

\hline
\multicolumn{3}{r}{\small\textit{Continued on next page\dots}} \\
\endfoot

\hline
\endlastfoot

\multicolumn{3}{|l|}{\textbf{Stage 1: Interactive Discovery and Pipeline Scaffolding (Google Antigravity)}} \\
\hline
Initial prompt & Antigravity & \textit{I am interested in designing a forecasting system for monthly Full Natural Flow (FNF) in California watersheds. The goal is to predict FNF 1 to 6 months into the future and create a model that may improve upon the Bulletin 120 forecasts. Discuss all relevant data sources for training and evaluating an operational forecasting model for this task.} \\
\hline
Research query & Antigravity & \textit{Search the scientific literature on seasonal streamflow forecasting in mountain watersheds. Identify actionable modeling ideas and physically motivated feature formulations that could improve model skill.} \\
\hline
Followup & Antigravity & \textit{Explore leading submissions to the Water Supply Forecast Rodeo competition (https://github.com/drivendataorg/water-supply-forecast-rodeo), which you just identified, and discuss how to adapt them to our task.} \\
\hline
Data investigation & Antigravity & \textit{Two data sources that are available in the FNF forecasting pipeline are not currently used by candidate models: daily\_ca\_sensors and snotel\_eom. Hypothesize how high-frequency end-of-month information from these datasets could be incorporated into the forecast system.} \\
\hline
Validation strategy & Antigravity & \textit{If I want to obtain a robust and accurate FNF forecasting model for California, and I use 1983--2020 for training and validation, what should be my training and validation splits? Discuss cross-validation strategies.} \\
\hline
Architectural guidance & Antigravity & \textit{Suggest at least three modeling architectures for FNF forecasting that could outperform operational B120 forecasts. Be mindful of overparameterization and overfitting risks.} \\
\hline
Feature engineering & Antigravity & \textit{Research candidate predictors that could improve the skill of model forecasts issued in spring.} \\
\hline
\multicolumn{3}{|l|}{\textbf{Stage 2: Autonomous MCTS Search and Code Mutation (ERA)}} \\
\hline
Code mutation suggestion & ERA & \textit{Experiment with the method used to select and aggregate station data as features, considering potentially missing data.} \\
\hline
Code mutation suggestion & ERA & \textit{Try reducing the complexity of the model without compromising on skill. Limit total parameter count given the small historical sample size.} \\
\hline
Code mutation suggestion & ERA & \textit{Experiment with different loss functions, aggregation weighting schemes, or auxiliary loss terms to better guide convergence.} \\
\hline
Code mutation suggestion & ERA & \textit{Selectively exclude predictors from the ensemble sub-models to explore further specialization.} \\
\hline
Persistent physical guideline & ERA & \textit{FNF forecasts must be non-negative.} \\
\hline
Persistent physical guideline & ERA & \textit{Impose monotone constraints between antecedent snow water content and FNF.} \\
\hline
\end{xltabular}
\endgroup

\clearpage

\begingroup
\small
\setlength{\tabcolsep}{4pt}
\renewcommand{\arraystretch}{0.98}
\begin{xltabular}{\textwidth}{|>{\raggedright\arraybackslash}p{2.45cm}|>{\raggedright\arraybackslash}X|>{\raggedright\arraybackslash}p{2.4cm}|>{\raggedright\arraybackslash}p{1.95cm}|>{\raggedright\arraybackslash}p{2.45cm}|>{\centering\arraybackslash}p{1.0cm}|}
\caption{Flowracle predictor features, including notation, physical description, observational data source, native spatial resolution (if any), temporal observation window, and mathematical formulation cross-reference. The final column indicates whether a positive monotonic constraint ($\partial \hat{y} / \partial x \ge 0$) is imposed during tree splitting ($\uparrow$), whether the feature enters unconstrained (---), or whether it is not a direct model input but a modulator or intermediate quantity (n/a).}\label{tab:features}\\
\hline
\textbf{Symbol} & \textbf{Feature Description} & \textbf{Data Source \& Res.} & \textbf{Window} & \textbf{Formulation} & \textbf{Mono.} \\
\hline
\endfirsthead

\multicolumn{6}{l}{\small\textit{Supplementary Table~\thetable, continued from previous page}} \\[4pt]
\hline
\textbf{Symbol} & \textbf{Feature Description} & \textbf{Data Source \& Res.} & \textbf{Window} & \textbf{Formulation} & \textbf{Mono.} \\
\hline
\endhead

\hline
\multicolumn{6}{r}{\small\textit{Continued on next page\dots}} \\
\endfoot

\hline
\endlastfoot

\multicolumn{6}{|l|}{\textbf{Calendar Harmonics}} \\
\hline
$\mathbf{h}_{\mathrm{iss}}(T)$ & Annual issuance calendar harmonics & Gregorian calendar & Month $T$ & Eq.~\eqref{eq:iss_harmonics} & --- \\
$\mathbf{e}_{\text{lead}}(T, l)$ & Lead horizon target harmonics and season flag & Gregorian calendar & Target month $m_l$ & Eq.~\eqref{eq:lead_encoding} & --- \\
\hline
\multicolumn{6}{|l|}{\textbf{Geographical and Topographical Coordinates}} \\
\hline
$\phi_s, \lambda_s$ & Basin outlet latitude and longitude & CDEC station metadata & Static & Eq.~\eqref{eq:sensor_dist} & --- \\
$z_s$ & Basin outlet / gauge elevation (m) & CDEC station metadata & Static & Eq.~\eqref{eq:sensor_dist} & --- \\
\hline
\multicolumn{6}{|l|}{\textbf{Antecedent Streamflow Memory (Full Natural Flow)}} \\
\hline
$\text{FNF}_{s, -2}$ & Two-month lagged streamflow & CDEC gauges & Month $T-2$ & Eq.~\eqref{eq:fnf_m2} ($F_s = 10^4$\,AF) & $\uparrow$ \\
$\overline{\text{FNF}}_s$ & Local catchment trailing runoff & CDEC gauges & Months $[T-6, T-2]$ & Eq.~\eqref{eq:fnf_mem} ($F_s = 10^4$\,AF) & $\uparrow$ \\
$\mathrm{FNF}_{s, \mathrm{nbr}, -2}$ & Two-month lagged neighbor streamflow & CDEC gauges & Month $T-2$ & Eq.~\eqref{eq:nbr_fnf_m2} ($F_s = 10^4$\,AF) & $\uparrow$ \\
$\overline{\mathrm{FNF}}_{s, \mathrm{nbr}}$ & Neighboring basins runoff & CDEC gauges & Months $[T-6, T-2]$ & Eq.~\eqref{eq:nbr_fnf_avg} ($F_s = 10^4$\,AF) & $\uparrow$ \\
\hline
\multicolumn{6}{|l|}{\textbf{Regional Hydroclimate and Thermal Forcing ($b \in \{\mathrm{N, MN, MS, S}\}$, $124^\circ\text{W}\text{--}115^\circ\text{W}$)}} \\
\hline
$\text{PPT}_b$ & Cumulative precipitation & PRISM (4\,km) & Months $[T-6, T-1]$ & Eq.~\eqref{eq:ppt} ($P_s = 60$\,mm) & $\uparrow$ \\
$\text{DD}_b$ & Positive degree-month melt sum & PRISM (4\,km) & Months $[T-6, T-1]$ & Eq.~\eqref{eq:dd} & --- \\
$\text{SWE}_b$ & Regional snow water equivalent & UA SWANN (4\,km) & Month $T-1$ & Eq.~\eqref{eq:swe} ($S_s = 450$\,mm) & $\uparrow$ \\
$R_b$ & Snowpack thermal ripeness index & PRISM + UA SWANN (4\,km) & Months $[T-6, T-1]$ & Eq.~\eqref{eq:ripeness} ($S_R = 100$\,mm) & --- \\
$\Delta\text{SWE}_b$ & Regional monthly SWE velocity & UA SWANN (4\,km) & Months $T-1$ vs. $T-2$ & Eq.~\eqref{eq:swe_vel} ($S_V = 50$\,mm) & $\uparrow$ \\
$\tau_b$ & Time elapsed since peak SWE & UA SWANN (4\,km) & Months $[T-6, T-1]$ & Eq.~\eqref{eq:time_since_peak} & --- \\
\hline
\multicolumn{6}{|l|}{\textbf{Catchment Telemetry and Dynamic Predictors ($\mathbf{f}_{\text{local}}(s, T)$)}} \\
\hline
$\text{PPT}_{\text{accum}, s}$ & Local cumulative precipitation & PRISM (4\,km) & Months $[T-6, T-1]$ & Eq.~\eqref{eq:local_ppt} ($P_s = 60$\,mm) & $\uparrow$ \\
$\text{SWE}_{\mathrm{ua}, s}$ & Local gridded snowpack & UA SWANN (4\,km) & Month $T-1$ & Eq.~\eqref{eq:local_ua_swe} ($S_s = 450$\,mm) & $\uparrow$ \\
$f_{\mathrm{snow}, s}$ & Local snow-to-precipitation ratio & PRISM + UA SWANN (4\,km) & Month $T-1$ vs. $[T-6, T-1]$ & Eq.~\eqref{eq:local_snow_frac} & $\uparrow$ \\
$\eta_{\text{prime}, s}$ & Primed runoff efficiency predictor & gridMET PDSI + PRISM (4\,km) & Months $[T-6, T-1]$ & Eqs.~\eqref{eq:primed_efficiency}--\eqref{eq:priming} & $\uparrow$ \\
$\text{SWE}_{\mathrm{snotel}, s}$ & Station-aggregated SNOTEL SWE & NRCS SNOTEL & Month $T-1$ & Eq.~\eqref{eq:snotel_swe} ($S_s = 450$\,mm) & $\uparrow$ \\
$\text{SWE}_{\mathrm{cdec}, s}$ & Station-aggregated CDEC SWE & CDEC snow pillows & Month $T-1$ ($\phi_s < 37^\circ\text{N}$) & Eq.~\eqref{eq:cdec_swe} ($S_s = 450$\,mm) & $\uparrow$ \\
$\text{SWE}_{\mathrm{eom, cdec}, s}$ & End-of-month CDEC pillow SWE & CDEC snow pillows & Final day ($T-1$; $\phi_s < 37^\circ\text{N}$) & Eq.~\eqref{eq:daily_cdec_swe} ($S_s = 450$\,mm) & $\uparrow$ \\
$\Delta\text{SWE}_{\mathrm{snotel}, s}$ & SNOTEL monthly rate of change & NRCS SNOTEL & Months $T-1$ vs. $T-2$ & Eq.~\eqref{eq:snotel_vel} ($S_V = 50$\,mm) & --- \\
$\Delta\text{SWE}_{\mathrm{cdec}, s}$ & CDEC monthly rate of change & CDEC snow pillows & Months $T-1$ vs. $T-2$ ($\phi_s < 37^\circ\text{N}$) & Eq.~\eqref{eq:cdec_vel} ($S_V = 50$\,mm) & --- \\
$\Delta_\text{eom}\text{SWE}_{\mathrm{snotel}, s}$ & Trailing-month SNOTEL SWE trend & NRCS SNOTEL & End-of-month vs. $T-1$ & Eq.~\eqref{eq:eom_diff} ($S_V = 50$\,mm) & $\uparrow$ \\
$\Delta_{\mathrm{daily}}\text{SWE}_{\mathrm{cdec}, s}$ & End-of-month CDEC SWE rate of change & CDEC snow pillows & Final 5 days ($T-1$; $\phi_s < 37^\circ\text{N}$) & Eq.~\eqref{eq:daily_cdec_vel} & $\uparrow$ \\
$f_{\mathrm{sn\_snow}, s}$ & SNOTEL snow-to-precipitation ratio & NRCS SNOTEL & Month $T-1$ / Water Year & Eq.~\eqref{eq:snotel_snow_frac} & $\uparrow$ \\
$V_{\mathrm{melt}, s}$ & Primed available snowmelt volume & NRCS SNOTEL + $\eta_{\text{prime}, s}$ & Month $T-1$ & Eq.~\eqref{eq:primed_melt_vol} & $\uparrow$ \\
$M_{\mathrm{onset}, s}$ & Primed melt-onset pulse & NRCS SNOTEL + $\eta_{\text{prime}, s}$ & Month $T-1$ ($T \in \{4, \dots, 8\}$) & Eq.~\eqref{eq:primed_melt_timing} & $\uparrow$ \\
$C_{\mathrm{storm}}$ & Cross-basin storm coherence & NRCS SNOTEL & Months $T-1$ vs. $T-2$ & Eq.~\eqref{eq:storm_coherence} & --- \\
$R_{\mathrm{bulk}, s}$ & State-to-basin snow ratio & NRCS SNOTEL & Months $T-1$ vs. $T-2$ & Eq.~\eqref{eq:storm_bulk_ratio} & --- \\
\hline
\multicolumn{6}{|l|}{\textbf{Drought Memory, Teleconnections, and Modulators}} \\
\hline
$\text{PDSI}_{T-1}$ & Palmer Drought Severity Index & gridMET (4\,km) & Month $T-1$ & Section~\ref{sec:methods_predictors}, Eq.~\eqref{eq:priming} & n/a \\
$\mathbf{h}_{\mathrm{mjo}}$ & Real-time Multivariate MJO harmonics & Australian BoM & Month $T-1$ (7-day trailing $\theta_{\mathrm{mjo}}$) & Eqs.~\eqref{eq:mjo_phase}--\eqref{eq:mjo_harmonics} & --- \\
$D_{\mathrm{late}, s}$ & Late storm damping (modulator) & NRCS SNOTEL & Months $[T-3, T-1]$ ($T = \text{April}$) & Eqs.~\eqref{eq:surge_ratio}--\eqref{eq:damped_daily_vel} ($\beta_{\mathrm{surge}} = 0.75$) & n/a \\
\hline
\end{xltabular}
\endgroup

\clearpage

\begingroup
\small
\setlength{\tabcolsep}{5pt}
\renewcommand{\arraystretch}{1.18}
\begin{xltabular}{\textwidth}{|>{\raggedright\arraybackslash}X|>{\centering\arraybackslash}p{2.8cm}|>{\centering\arraybackslash}p{2.8cm}|>{\centering\arraybackslash}p{3.2cm}|}
\caption{Hyperparameters and feature specialization for the three XGBoost sub-models comprising Flowracle. Shared specifications include the \texttt{hist} tree building algorithm, the loss objective (\texttt{reg:quantileerror}) across 13 evaluation quantiles $\tau \in \{0.01, 0.05, 0.10, \dots, 0.90, 0.95, 0.99\}$, and positive monotonic constraints on physical moisture and snowpack features.}\label{tab:submodels}\\
\hline
\textbf{Hyperparameter / Attribute} & \textbf{Anchor} ($\mathcal{M}_{\text{anchor}}$) & \textbf{Core} ($\mathcal{M}_{\text{core}}$) & \textbf{Anomaly} ($\mathcal{M}_{\text{anom}}$) \\
\hline
\endfirsthead

\multicolumn{4}{l}{\small\textit{Supplementary Table~\thetable, continued from previous page}} \\[4pt]
\hline
\textbf{Hyperparameter / Attribute} & \textbf{Anchor} ($\mathcal{M}_{\text{anchor}}$) & \textbf{Core} ($\mathcal{M}_{\text{core}}$) & \textbf{Anomaly} ($\mathcal{M}_{\text{anom}}$) \\
\hline
\endhead

\hline
\multicolumn{4}{r}{\small\textit{Continued on next page\dots}} \\
\endfoot

\hline
\endlastfoot

\multicolumn{4}{|l|}{\textbf{Target Formulation and Tree Architecture}} \\
\hline
Target optimization variable & Raw log-FNF ($y$) & Raw log-FNF ($y$) & Climatological residual ($y_{\text{resid}}$) \\
\hline
Max tree split depth ($d_{\max}$) & $5$ & $9$ & $6$ \\
\hline
Number of boosting trees ($n$) & $4{,}000$ & $7{,}000$ & $6{,}000$ \\
\hline
Shrinkage learning rate ($\eta$) & $0.008$ & $0.005$ & $0.005$ \\
\hline
\multicolumn{4}{|l|}{\textbf{Regularization and Stochastic Sampling}} \\
\hline
Base L2 regularization penalty ($\lambda_0$) & $780$ & $1{,}430$ & $750$ \\
\hline
Minimum child weight (Hessian sum per leaf) & $80$ & $135$ & $95$ \\
\hline
Stochastic row bagging ratio (\texttt{subsample}) & $0.85$ & $0.85$ & $0.85$ \\
\hline
Column feature subsampling ratio (\texttt{colsample\_bytree}) & $0.55$ & $0.55$ & $0.55$ \\
\hline
\multicolumn{4}{|l|}{\textbf{Feature Selection and Domain Specialization}} \\
\hline
Basin outlet coordinates & Included & Included & Excluded \\
\hline
MJO (amplitude and phase) & Excluded & Excluded & Included \\
\hline
Time since peak SWE ($\tau_b$) & Excluded & Included & Excluded \\
\hline
Snowpack thermal ripeness index ($R_b$) & Excluded & Included & Included \\
\hline
Neighbor trailing mean runoff ($\overline{\mathrm{FNF}}_{s,\mathrm{nbr}}$) & Included & Excluded & Included \\
\hline
April--July target season indicator ($\mathbb{1}_{\mathrm{AMJJ}, l}$) & Included & Included & Excluded \\
\hline
Retained predictors (total of $55$) & $44$ & $51$ & $47$ \\
\hline
\end{xltabular}
\endgroup

\clearpage

\begin{figure}[tbh!]
\centering
\includegraphics[width=\textwidth]{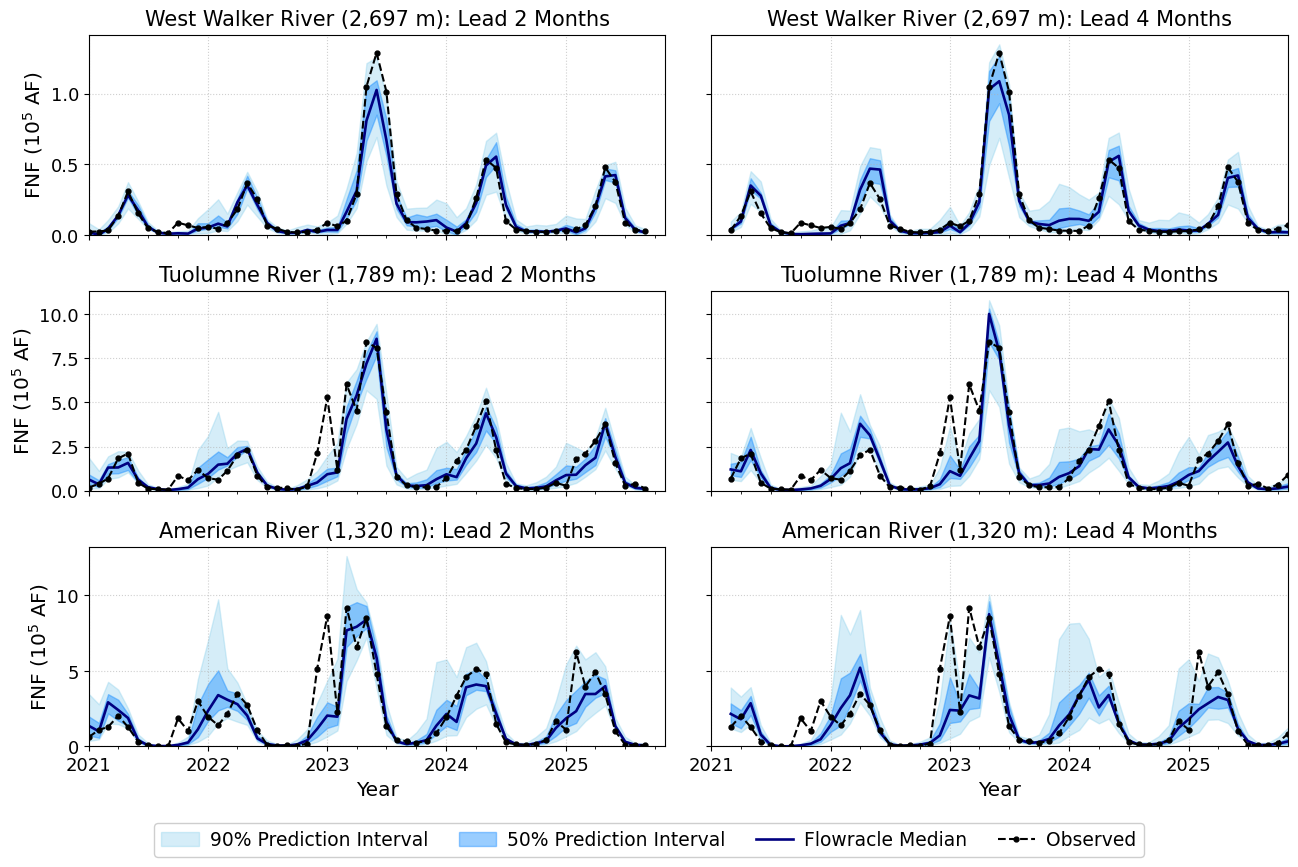}
\caption{Probabilistic forecast time series across a Sierra Nevada elevation transect for 2- and 4-month lead times (2021--2025). Panels display high-elevation (West Walker River), mid-elevation (Tuolumne River), and low-elevation (American River) catchments, with 90\% and 50\% prediction intervals, median Flowracle forecasts, and observed flow.}\label{fig:si_fan_charts}
\end{figure}

\begin{figure}[tbh!]
\centering
\includegraphics[width=\textwidth]{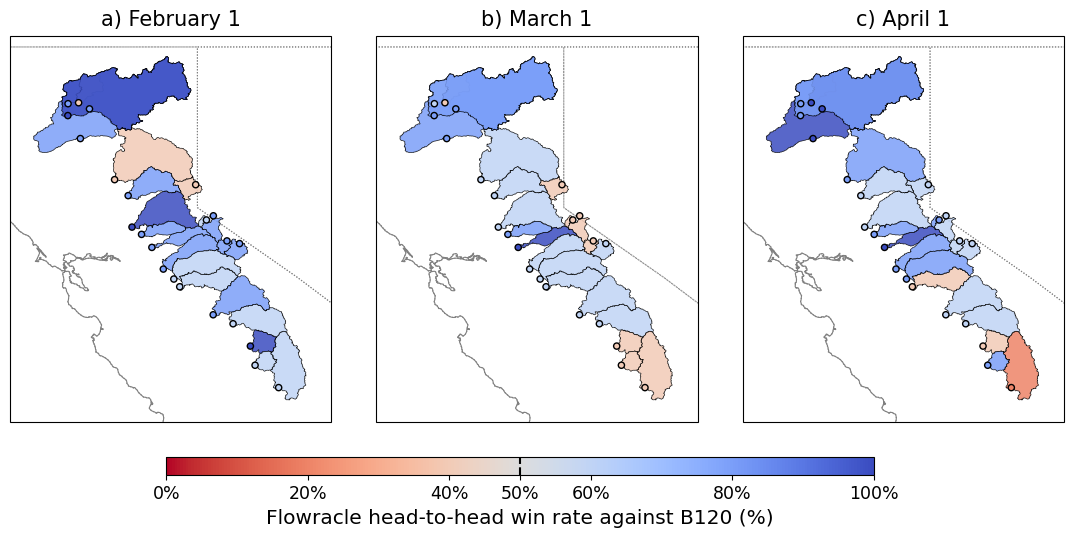}
\caption{Forecast win rate against B120 (2021--2025) by issuance month across all 23 snow-fed basins: \textbf{(a)} February 1, \textbf{(b)} March 1, and \textbf{(c)} April 1. Flowracle wins 72.2\%, 58.3\%, and 70.4\% of basin-year trials statewide in February, March, and April, respectively.}\label{fig:si_spatial_win_rates}
\end{figure}

\end{document}